\documentclass[aps,prd,english,preprintnumbers,nofootinbib,floatfix,twocolumn,10pt]{revtex4-1}

\usepackage{amsfonts,amsmath,amssymb}
\usepackage{graphicx,epsfig}
\usepackage{float}
\usepackage{subfig}
\usepackage{subfloat}
\usepackage[utf8]{inputenc}
\usepackage{hyperref}
\usepackage{babel}

\usepackage[font=footnotesize,labelfont=rm,justification=justified]{caption}

\newcommand\be{\begin{equation}}
\newcommand\ee{\end{equation}}
\newcommand\bea{\begin{eqnarray}}
\newcommand\eea{\end{eqnarray}}
\newcommand\bx{{\bar x}}

\begin{document}

\def\rhoo{\rho_{_0}\!} 
\def\rhooo{\rho_{_{0,0}}\!} 

\begin{flushright}
\phantom{
{\tt arXiv:1404.$\_\_\_\_$}
}
\end{flushright}

{\flushleft\vskip-1.4cm\vbox{\includegraphics[width=1.15in]{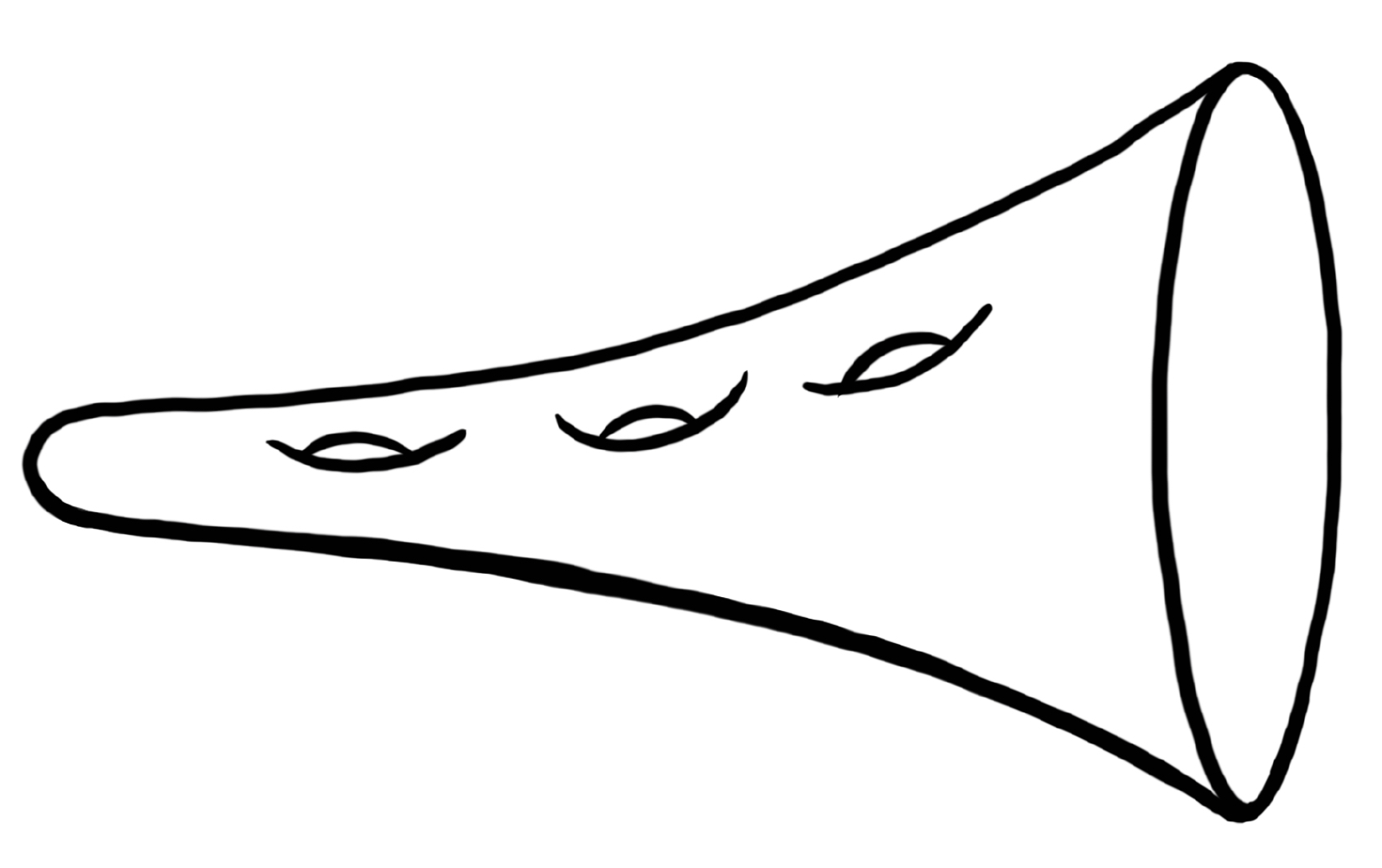}}}

\title{JT Supergravity, Minimal Strings, and  Matrix Models}
\author{Clifford V. Johnson}
\email{johnson1@usc.edu}
\affiliation{Department of Physics and Astronomy\\ University of
Southern California \\
 Los Angeles, CA 90089-0484, U.S.A.}


\begin{abstract}
It is proposed that a family of Jackiw--Teitelboim supergravites, recently discussed in connection with  matrix models by Stanford and Witten,  can be  given a complete definition, to all orders in the topological expansion 
and beyond, in terms of a specific combination of minimal string theories.  This construction defines   non--perturbative physics for the supergravity that    is well--defined and stable.  The minimal models come from  double--scaled complex matrix models and  correspond to the cases $(2\Gamma{+}1,2)$  in the  Altland--Zirnbauer $(\boldsymbol{\alpha},\boldsymbol{\beta})$ classification of random matrix ensembles, where $\Gamma$ is a parameter. A central role is played by a non--linear ``string equation'' that 
 naturally incorporates~$\Gamma$, usually taken to be an integer, counting {\it e.g.,} D--branes in the minimal models. 
 Here, half--integer~$\Gamma$ also has an interpretation. In fact,~$\Gamma{=}{\pm}\frac12$ yields  the cases $(0,2)$ and $(2,2)$  that were shown  by Stanford and Witten to  have very special properties.  These features are manifest in this definition  because the relevant  solutions of the string equation have special properties for~$\Gamma{=}{\pm}\frac12$. Additional special features for other half--integer~$\Gamma$  suggest new surprises in the supergravity models.   
\end{abstract}


\maketitle

\section{Introduction}
\label{sec:introduction}

Recently, there has been  renewed  interest in Jackiw--Teitelboim (JT) gravity~\cite{Jackiw:1984je,Teitelboim:1983ux}, a two dimensional  theory of gravity that emerges in  various physical contexts, such as the study of the near--horizon dynamics of nearly extreme black holes~\cite{Fabbri:2000xh,Nayak:2018qej}, or as a partial gravitational dual of certain one--dimensional quantum mechanical systems pertinent to studies of condensed matter and quantum chaos. (For  reviews, see refs.~\cite{Grumiller:2002nm,Sarosi:2017ykf}.) In its own right, it is an  arena for  further developing understanding of the interplay between quantum mechanics, geometry, and topology.

The JT gravity partition function $Z(\beta)$ can be written  (in a Euclidean presentation, where $\beta$ is the period of compact time)  as a topological expansion summing contributions from constant negative curvature surfaces  of genus $g$ (the number of handles) with a boundary of fixed length~$\beta$.  

The dynamics of the theory lives on the boundary, and it has a Schwarzian action~\cite{Almheiri:2014cka,Jensen:2016pah,Maldacena:2016upp}. The leading contribution, which comes from the $g{=}0$ (disc) topology, gives a result $Z_0(\beta)$ which can be written~\cite{Cotler:2016fpe}:
\be
\label{eq:density}
Z_0(\beta) = e^{S_0}\!\!\int \!\!dE\, \rhoo(E) e^{-\beta E}\ ,
\ee
where $\rhoo(E)$ is a spectral density function. Here,~$S_0$ is a constant proportional to $1/G$, where $G$ is the Newton constant of the 2D gravity. (In fact, $S_0$ is  the leading black hole entropy, if approaching this model from a near--horizon dynamics perspective.) Here, it will be useful to define a parameter $\hbar{=}e^{-S_0}$, since it will naturally appear as a Planck constant in an important associated problem to be described shortly. The result for the disc spectral density is~\cite{Maldacena:2016hyu}:
\be
\label{eq:spectral-JT}
\rhoo(E)=\frac{\gamma\sinh(2\pi\sqrt{2\gamma E})}{2\pi^2\hbar}\ ,\qquad{\rm (JT)}
\ee
where  $\hbar^{-1}$ will be absorbed into its definition henceforth. The value of $\gamma$ determines the  units when relating a coupling in the Schwarzian  to  $\beta$. Here $\gamma{=}\frac12$ will be chosen.

JT gravity emerges~\cite{Almheiri:2014cka,Jensen:2016pah,Maldacena:2016upp,Engelsoy:2016xyb} (at low energy)  as a  gravitational dual of certain 1D quantum systems that exhibit chaos, such as the    Sachdev--Ye--Kitaev (SYK) model~\cite{Sachdev:1992fk,Kitaev:talks,Maldacena:2016hyu}, and various features of the chaotic dynamics suggested~\cite{Garcia-Garcia:2016mno,Cotler:2016fpe,Saad:2018bqo} a relation to models of large~$N$ random matrix models. Then Saad, Shenker and Stanford showed~\cite{Saad:2019lba} that the entire topological expansion for JT gravity can be captured by  an Hermitian matrix model in a certain ``double--scaling'' limit~\cite{Brezin:1990rb,Douglas:1990ve,Gross:1990vs,Gross:1990aw}.   The double--scaled~$1/N$ expansion of the model gives a genus expansion in 2D surfaces~\cite{'tHooft:1973jz,Brezin:1978sv}, and has its contributions at higher genus fully determined by a family of recursion relations~\cite{Eynard:2004mh,Mirzakhani:2006fta,Eynard:2007fi,Eynard:2007kz} seeded by the disc spectral density $\rhoo(E)$, which was shown~\cite{Saad:2019lba} to precisely match analogous features of JT gravity.

The original double--scaled matrix models~\cite{Brezin:1990rb,Douglas:1990ve,Gross:1990vs,Gross:1990aw} were used to define the sum over random surfaces that were the string world--sheets in what have now come to be called ``minimal string theories''~\cite{Seiberg:2004at}. While not itself a minimal string theory, JT gravity shares a number of features with them, and ref.~\cite{Saad:2019lba} suggested that JT gravity can be thought of as an infinite order limit of minimal string theories. An (apparently) alternative picture, suggested in ref.~\cite{Okuyama:2019xbv}  and expanded upon in ref.~\cite{Johnson:2019eik}, (see also the recent refs.~\cite{Betzios:2020nry,Okuyama:2020ncd}), is that JT gravity can be defined as a special interpolating flow among an infinite set of minimal string models. In fact, it will be proposed in section~\ref{sec:minimal-models} 
that the two suggestions are complementary, the latter being a refinement of the former.

The approach of constructing JT gravity out of minimal string models has certain advantages, as will be further demonstrated in this paper.  The minimal models (arising from one--cut double--scaled matrix models) have a great deal of their physics readily accessible through an associated 1D problem, with Hamiltonian 
\be
\label{eq:schrodinger}
{\cal H}=-\hbar^2\frac{\partial^2}{\partial x^2}+u(x)\ , 
\ee
 where the potential $u(x)$ satisfies a non--linear ordinary differential equation (ODE) called a ``string equation''. This ODE supplies both perturbative and non--perturbative information for $u(x)$, and the spectral density  of this system coincides with the spectral density of the double--scaled matrix model. Knowledge of the properties of the underlying string equation, especially non--perturbative features, can be used to infer properties of the JT model that are not apparent (or simply inaccessible) in the perturbative/recursive approach. For example, ref.~\cite{Johnson:2019eik} used this insight to formulate  a non--perturbative low energy completion of the JT matrix model of ref.~\cite{Saad:2019lba} that is free of the instabilities of the  definition based on Hermitian matrix models.

In fact, the double--scaled matrix model description  extends to other kinds of JT gravity. Various JT gravity and JT supergravity models were classified by Stanford and Witten~\cite{Stanford:2019vob} in terms of the underlying distinct random matrix ensembles available. They are either from the three $\beta$--ensembles {\it \`a la} Dyson--Wigner, or the seven $(\boldsymbol{\alpha},\boldsymbol{\beta})$  ensembles in the Altland--Zirnbauer taxonomy~\cite{Altland:1997zz}. In a series of non--trivial computations, the properties of the various matrix model recursion relations were checked against the corresponding JT gravity--type computations, and found to support the correspondence. Various other (mostly perturbative) features were uncovered in that work as well. 

It is therefore natural to wonder if this wider class of JT gravities and their matrix models can be constructed out of appropriate kinds of minimal string models. Perhaps this could help to clarify their properties, or supply information about the non--perturbative sector, since the recursive methods of refs.~\cite{Saad:2019lba,Stanford:2019vob} are perturbative.

This paper will show that such constructions are possible, at least for some of the models. The focus will be directly on a set of JT supergravity theories (with and without time reversal symmetry)  that were discussed in ref.~\cite{Stanford:2019vob}, which is of relevance to ${\cal N}{=}1$ supersymmetric generalizations of the Schwarzian dynamics that arose in ordinary JT gravity, or the SYK model\cite{Fu:2016vas,Li:2017hdt,Kanazawa:2017dpd,Forste:2017kwy,Murugan:2017eto,Sun:2019yqp}. 
The disc spectral density of these models is~\cite{Stanford:2017thb,Stanford:2019vob}: 
\be
\label{eq:spectral-SJT}
\rhoo(E)=\frac{\cosh(2\pi\sqrt{E})}{\pi\hbar\sqrt{ E}}=\frac{\rhoo^{\rm SJT}}{\sqrt{2}}\ ,\qquad{\rm SJT}
\ee
where the factor of $\sqrt{2}$ gives the appropriate normalization for comparing the pertinent matrix ensembles to the supergravity path integral.

Before proceeding, it is  prudent to enlarge the notation slightly.
So far, the ``0'' subscript on the partition function and the corresponding density means that the result is at disc order in the genus expansion. In this broader class of theories, because of the possible inclusion of time--reversal symmetry, spacetime can be non--orientable, and so crosscaps should be included in the topological sum. Quantities such as the spectral density function will therefore be written perturbatively as:
\be
\rho(E)=\sum_{g,b,c}\hbar^{2g+c-1}\rho_{g,c}(E)\ ,
\ee
where $g$ is the number of handles of the surface, and $c$ the number of crosscaps. So in this notation, $\rhoo(E)\equiv\rhooo(E)$. Henceforth, the powers of $\hbar$ will be explicitly included in the density, as already done in {\it e.g.,} equations~(\ref{eq:spectral-JT}) and~(\ref{eq:spectral-SJT}).
\begin{figure}[h]
\centering
\includegraphics[width=0.44\textwidth]{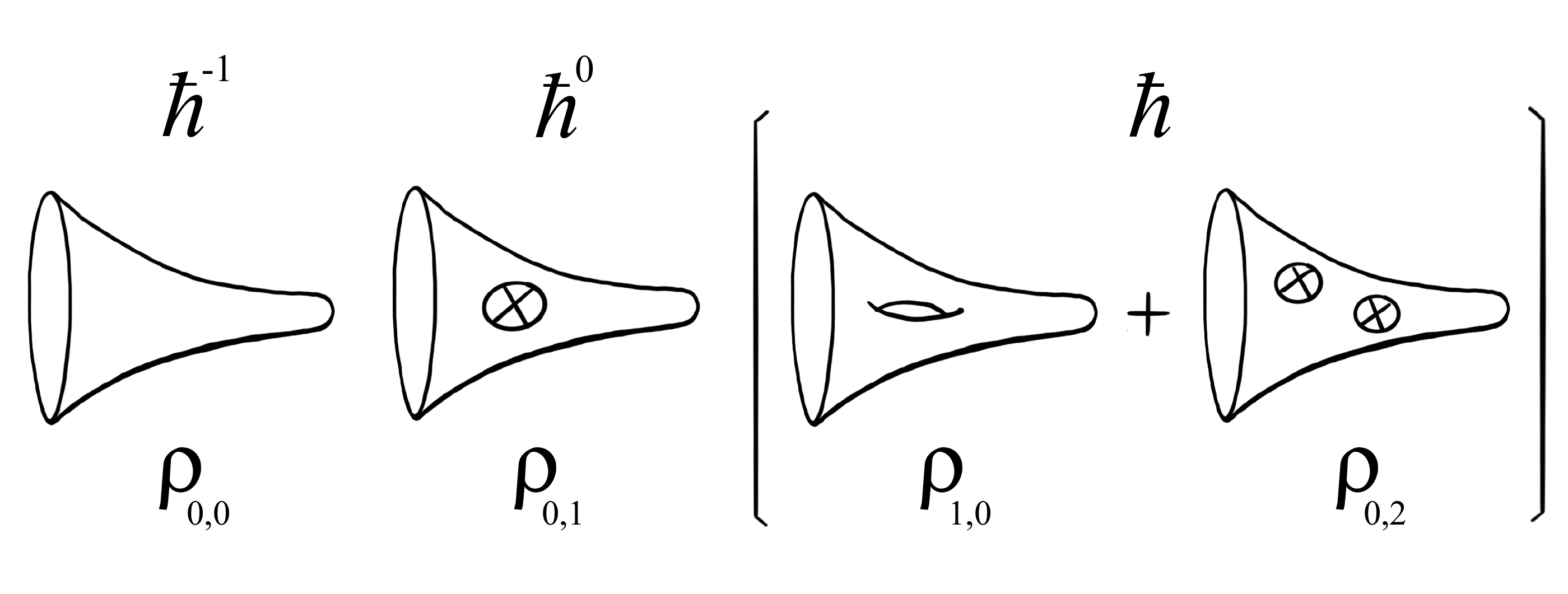}
\caption{\label{fig:orders} Examples of perturbative contributions. A cross represents a crosscap insertion making a non--orientable surface.}
\end{figure}
See figure~\ref{fig:orders} for the first three orders.

The  core specific JT supergravity models in question are the $(\boldsymbol{\alpha},\boldsymbol{\beta}){=}(\{0,1,2\} ,2)$  in the Altland--Zirnbauer classification. Many perturbative aspects of these models were constructed and exhibited in ref.~\cite{Stanford:2019vob}, including several important peculiarities, such as the vanishing of perturbative contributions to $\rhoo$ beyond the disc for the cases $\alpha{\in}\{0,2\}$. (As will become clear later, however, a broader class of models, $(2\Gamma{+}1,2)$, will also be accessible {\it via} the methods of this paper too, and results for those will also be presented.)

The present work will supply a description of these supergravities in terms of an infinite interpolating set of  minimal string models (of type~0A~\cite{Morris:1990bw,Morris:1991cq,Dalley:1992qg,Dalley:1992vr,Dalley:1992yi,Klebanov:2003wg}), {\it i.e.,} using the same ingredients presented in ref.~\cite{Johnson:2019eik}, but combining them  differently (see sections~\ref{sec:minimal-models} and~\ref{sec:SJT-from-minimal}). This will yield the spectral density~(\ref{eq:spectral-SJT}) at disc level, in a way that generalizes the case for ordinary JT in an interesting (and suggestive) manner.

The approach used (the aforementioned string equations, and  associated Hamiltonian problem~(\ref{eq:schrodinger})---see section~\ref{sec:string-equations} for introduction) will be quite complementary to the approaches of refs.~\cite{Saad:2019lba,Stanford:2019vob}, and will have the advantage of making more manifest certain perturbative features of the models. For example, the type~0A minimal strings ({\it via} their defining string equation) have {\it just} the right properties (see section~\ref{sec:type-0A}) needed to yield the key leading (universal)  $1/\sqrt{E}$ dependence at low energy. Their (fractional) power law rise with $E$ for higher energies collectively contribute to the overall exponential rise of the full model.  As a further example, the  vanishing (observed in ref.~\cite{Stanford:2019vob}) of all perturbative contributions (beyond the leading terms) for $\boldsymbol{\alpha}{\in}\{0,2\}$ will follow straightforwardly in this formalism, due to certain special properties of the string equations. (This is discussed in section~\ref{sec:large-positive-x})
 
 
Moreover, the present approach allows  a clear formulation of the {\it full} non--perturbative physics of the models that reduces to the disc result~(\ref{eq:spectral-SJT}) for all $E$.  This supplements and extends to all energies the non--perturbative aspects that were  touched upon   in ref.~\cite{Stanford:2019vob}, which applied mostly to the low energy regime.  Model behaviours will  be displayed in section~\ref{sec:non-perturbative}. 

Another appealing feature of this paper's approach is that the formulation allows for a larger family of models ---$(2\Gamma{+}1,2)$ in the classification scheme---to be cast into a single framework. The parameter $\Gamma$ is likely to be identified with the Stanford--Witten parameter $\nu$, counting ``Ramond punctures''~\cite{Stanford:2019vob}. (See more discussion of this in sections~\ref{sec:conclusions} and~\ref{sec:large-negative-x}.) Again, certain perturbative features become manifest here. For example, ref.~\cite{Stanford:2019vob}'s observation that only even numbers of punctures are allowed becomes a straightforward manifest feature of the formulation. The special cases $\boldsymbol{\alpha}{\in}\{0,2\}$ correspond to $\Gamma{=}{\pm}\frac12$, which is a special point of all type~0A solutions of the string equation.  However, the string equation shows that there are other special features for more general half--integer $\Gamma$, suggesting that there are new aspects of these JT supergravities to be discovered in these cases. The present formulation supplies full non--perturbative physics here too. See sections~\ref{sec:large-positive-x} and~\ref{sec:non-perturbative}.

Section~\ref{sec:conclusions} presents some closing remarks and ideas for further exploration.

\section{Spectral Densities from\\ \hskip 0.6cm Minimal Strings: Disc level}
\label{sec:minimal-models}

The JT gravity (and the JT supergravity) partition function  is structurally the same as  a ``macroscopic loop'' expectation value in the old  double--scaled matrix model language~\cite{Banks:1990df,Ginsparg:1993is}, involving the trace of an effective one dimensional Hamiltonian~(\ref{eq:schrodinger}),  that arises naturally from the matrix model after double scaling:
\be
\label{eq:JT-definition}
Z(\beta)= \int_{-\infty}^\mu\!\! dx\, \langle x| e^{-\beta{\cal H}({\hat p},{\hat x})} |x\rangle\ ,
\ee
where the upper limit on the $x$--integration, $\mu$,  will be discussed below.
Instead of fixed loop length $\ell$, the problem describes fixed inverse temperature $\beta$, which is the length of the boundary of the nearly--AdS$_2$ spacetime here. As a quantum mechanics problem, to examine the disc--level physics all that is needed is to work at leading  order in~$\hbar$. Denoting   the leading/classical piece of the potential  as $u_0(x){=}\lim_{\hbar\to0}(u(x))$,  inserting of a complete set of momentum states and using the following normalization for the wavefunction: $\langle x|p\rangle{=}e^{ipx}/\sqrt{2\pi\hbar}$, write
\bea
Z_0(\beta)
&=&\int_{-\infty}^\mu \!\!dx \int_{-\infty}^{+\infty}\frac{dp}{2\pi\hbar} e^{-\beta[p^2+u_0(x)]} \nonumber \\
&=&\frac{1}{2\hbar\sqrt{\pi\beta}}\int_{-\infty}^\mu \!\!dx\, e^{-\beta u_0(x)}\nonumber\\
&=& \frac{1}{2\pi\hbar}\sqrt{\frac{\pi}{\beta}}\int_{0}^\infty du_0 f(u_0)  e^{-\beta u_0}\ ,
\eea
where $f(u_0){=}{-}\partial x/\partial u_0$. The fact that $u_0(\mu){=}0$ was used, which will be confirmed below. The last integral  can be written as:
\bea
Z_0(\beta)&=&\int_{0}^\infty dE \int_{0}^E\frac{f(u_0)}{\sqrt{E-u_0}}\frac{du_0}{2\pi\hbar}e^{-\beta E}\ ,\nonumber\\
&=& \int_{0}^\infty\! dE\, \rhoo(E)\, e^{-\beta E} \ , 
\eea
where 
\be
\label{eq:spectral-deconstruction}
\rhoo(E)=\frac{1}{2\pi\hbar} \int_{0}^E\frac{f(u_0) }{\sqrt{E-u_0}}\, du_0\ . 
\ee

So $\rhoo(E)$ is determined if  $u_0(x)$ is known, since it defines the $f(u_0)$ kernel of the integral transform in equation~(\ref{eq:spectral-deconstruction}).   The lower limit  is $u_0{=}0$, and can be seen to mark the $E{=}0$ end of the classical spectral density in this construction. A popular and important case is the simple (``Airy'') potential $u(x){=}{-}x$. So $u_0(x){=}{-}x$, and $f(u_0){=}1$. This readily yields $\rhoo(E){=}E^{1/2}/\pi\hbar$. This is the double--scaled limit of the famous Wigner semi--circle law for a Gaussian Hermitian matrix model.

\begin{figure}[h]
\centering
\includegraphics[width=0.45\textwidth]{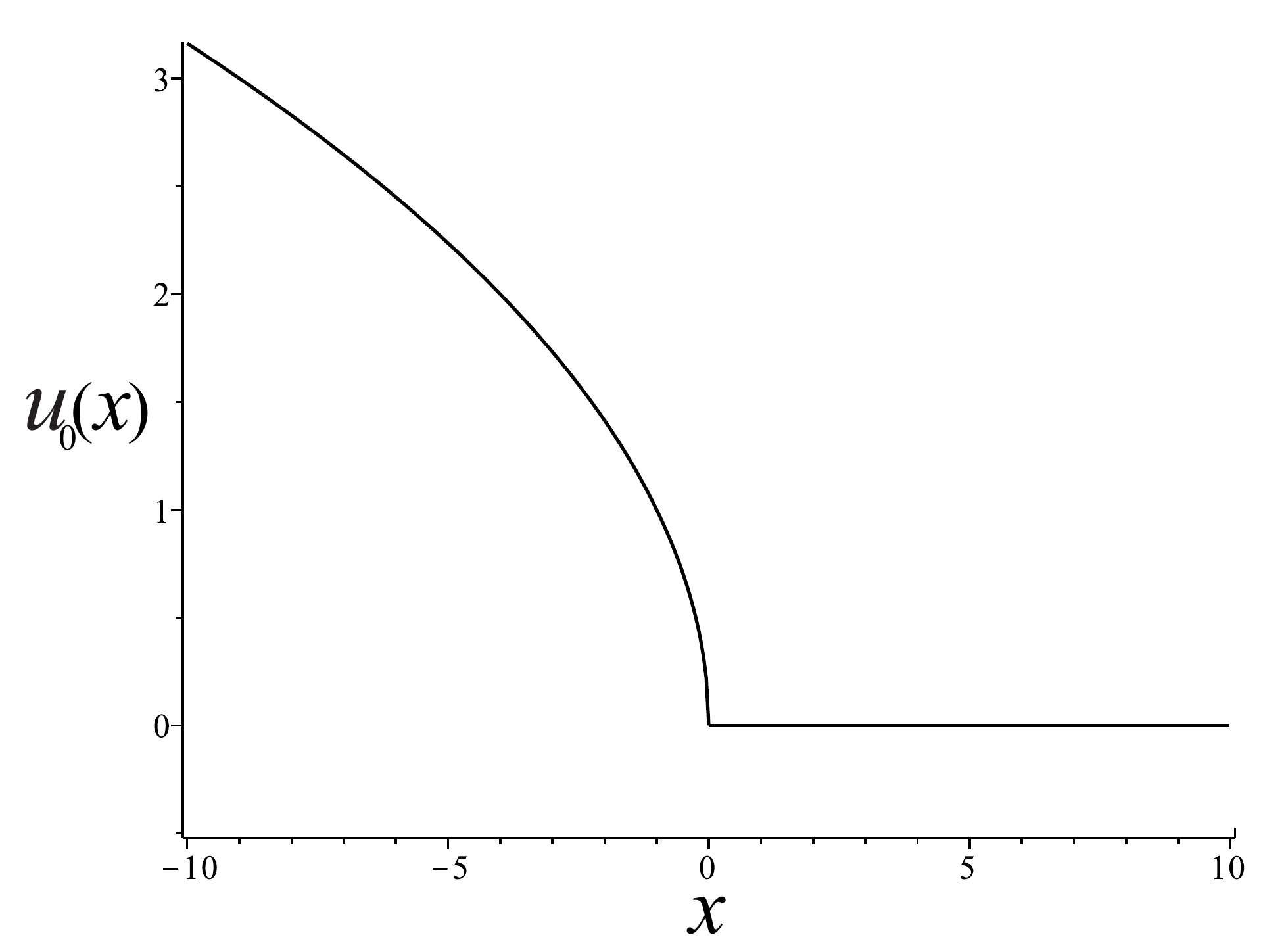}
\caption{\label{fig:leading-potential} Features of the  leading potential for the minimal models. For $x{<}\mu$, $u_0(x){=}(-x)^{1/k}$ ($k{=}2$ plotted here). This is common to both the bosonic and the type~0A  minimal models. For $x{>}\mu$, $u_0(x){=}0$, a key feature of the type~0A minimal models.  (Figure~\ref{fig:potential2} shows the full $u(x)$ for $k{=}2$ type~0A.)}
\end{figure}

Of interest will be the behaviours indexed by integer~$k$: $u_0{=}(-x)^{1/k}$ for large $x{<}0$, a well--known leading behaviour for certain classes of minimal string models ({\it i.e.,} the ``multicritical'' behaviour in the old double--scaling language~\cite{Kazakov:1989bc}). (See the left region of figure~\ref{fig:leading-potential}.) Simple scaling of equation~(\ref{eq:spectral-deconstruction}) shows that this will yield a spectral density  $\rhoo{=}C_k E^{k-\frac12}/2\pi\hbar$, but the numerical coefficient~$C_k$ will be important in what follows. A bit of work shows that, with $z{=}u_0/E$:
\bea
\label{eq:Ck}
C_k&=&k\int_0^1\frac{z^{k-1}dz }{\sqrt{1-z}}=2k\int_0^{\pi/2}(\sin\theta)^{2k-1}d\theta \nonumber\\
&=& \frac{2^{2k-1}((k-1)!)^2k}{(2k-1)!}\ ,
\eea
which can be proven by {\it e.g.,} expressing $(\sin\theta)^n$, for odd~$n$, in terms of sums of terms involving $\sin(n\theta), \sin((n{-}2)\theta),$ {\it etc}. The coefficients are sums of those in the binomial expansion.

Now it is time to build more complicated spectral densities relevant to JT and super JT. For any $k$, the potential $u_0{=}(-x)^{1/k}$ increases as $x{\to}{-}\infty$, and so this form  will dominate the large $E$ behaviour. A general minimal model (in this class)  has this defining equation for~$u_0$,  as $x{\to}{-}\infty$:
\be
\label{eq:string-equation-sphere}
\sum_{k=1}^\infty t_k u_0^k=- x\ ,
\ee
where $t_k$  is the coupling that turns on the $k$th model. This equation is to be thought of as an interpolating flow connecting all the models. Such an equation is the leading piece of what was called a ``string equation'' in the older matrix model literature. (The full non--linear string equations will be discussed in section~\ref{sec:string-equations}.) So, in preparation for working with equation~(\ref{eq:spectral-deconstruction}):
\be
\label{eq:fsum}
f(u_0)=-\frac{\partial x}{\partial u_0}=\sum_{k=1}^\infty kt_k u_0^{k-1}\ .
\ee
The scheme to move forward with here is that any Schwarzian--type disc level spectral density $\rhoo(E)$ that has a series expansion in powers of the form $E^{k-\frac12}$ can be reconstructed from a potential $u_0(x)$ that can be deduced by using the ingredients above. It amounts to a specific formula for the $t_k$, determining the particular combination of minimal models to be used to build the JT gravity model. For example, in the case of the ordinary JT model, with the disc spectral density given in equation~(\ref{eq:spectral-JT}):

 \bea
 \label{eq:JT-deconstructed}
 \rhoo(E)&=&\frac{1}{4\pi^2 \hbar}\sum_{k=1}^\infty\frac{(2\pi\sqrt{E})^{2k-1}}{(2k-1)!}\qquad ({\rm JT})\nonumber\\
 &=&\frac{1}{4\pi^2 \hbar}\int_0^E  \sum_{k=1}^\infty \frac{2^{2k-1}\pi^{2k-1}}{(2k-1)!}\frac{k}{C_k} u_0^{k-1}  \frac{du_0}{\sqrt{E-u_0}} \nonumber\\
 &=&\frac{1}{4\pi \hbar}\int_0^E  \sum_{k=1}^\infty \frac{\pi^{2k-2}u_0^{k-1}}{(k-1)!^2}   \frac{du_0}{\sqrt{E-u_0}} \nonumber\\
 &=&\frac{1}{4\pi \hbar}\int_0^E  \sum_{k=1}^\infty \frac{(\pi\sqrt{u_0})^{2k-2}}{(k-1)!^2}   \frac{du_0}{\sqrt{E-u_0}} \nonumber\\
 &=& \frac{1}{2\pi \hbar}\int_0^E  \frac{I_0(2\pi\sqrt{u_0})du_0}{2\sqrt{E-u_0}} \ ,
 \eea
 where $I_0(s)$ is the zeroth modified Bessel function of the first kind. From the above, it is clear that  from equations~(\ref{eq:spectral-deconstruction}) and~(\ref{eq:fsum}) that the $t_k$ are:
\be
\label{eq:tk-recipe-JT}
t_k=\frac{\pi^{2k-2}}{2k!(k-1)!}\ ,
\ee
as shown in ref.~\cite{Okuyama:2019xbv} (with a different normalization), and used in ref.~\cite{Johnson:2019eik}, as described in the Introduction.

It is worth pausing to reflect on the meaning of this. First, and most importantly, note that the simple additive structure   in equation~(\ref{eq:string-equation-sphere}) is  deceptively simple. For each minimal model the full string equation (see section~\ref{sec:string-equations}) will be highly non--linear in $u(x)$ and its derivatives, and the additive structure will couple together those non--linear equations. So the  resulting solution $u(x)$ is {\it not} a simple sum of the behaviour of the $u(x)$s from the individual models. Remarkably however, $u(x)$'s evolution as a function of the $t_k$ is  described by the $k$th (integrable) KdV flow equation~\cite{Douglas:1990dd,Banks:1990df}, which takes the form:
\be
\label{eq:KdV-flows}
\frac{\partial u}{\partial t_k}\propto\frac{\partial}{\partial x}{\tilde R}_{k+1}[u]\ ,
\ee
where the ${\tilde R}_k$ are polynomials in $u(x)$ and its derivatives (examples are listed in equation~(\ref{eq:gelfand-dikii})) and  the $t_k$ are the times of the integrable flows. From a conformal  field theory perspective, for {\it e.g.,} the bosonic $(2,2k{-}1)$ minimal models (similar remarks hold for the type~0A models discussed later),  the $k$th model has $k{-}1$ operators,~${\cal O}_l$, ($l{=}1,2\cdots,k{-}1$), and deformation with  $t_l$ is equivalent to turning on the operator ${\cal O}_l$. 
\begin{figure}[h]
\centering
\includegraphics[width=0.48\textwidth]{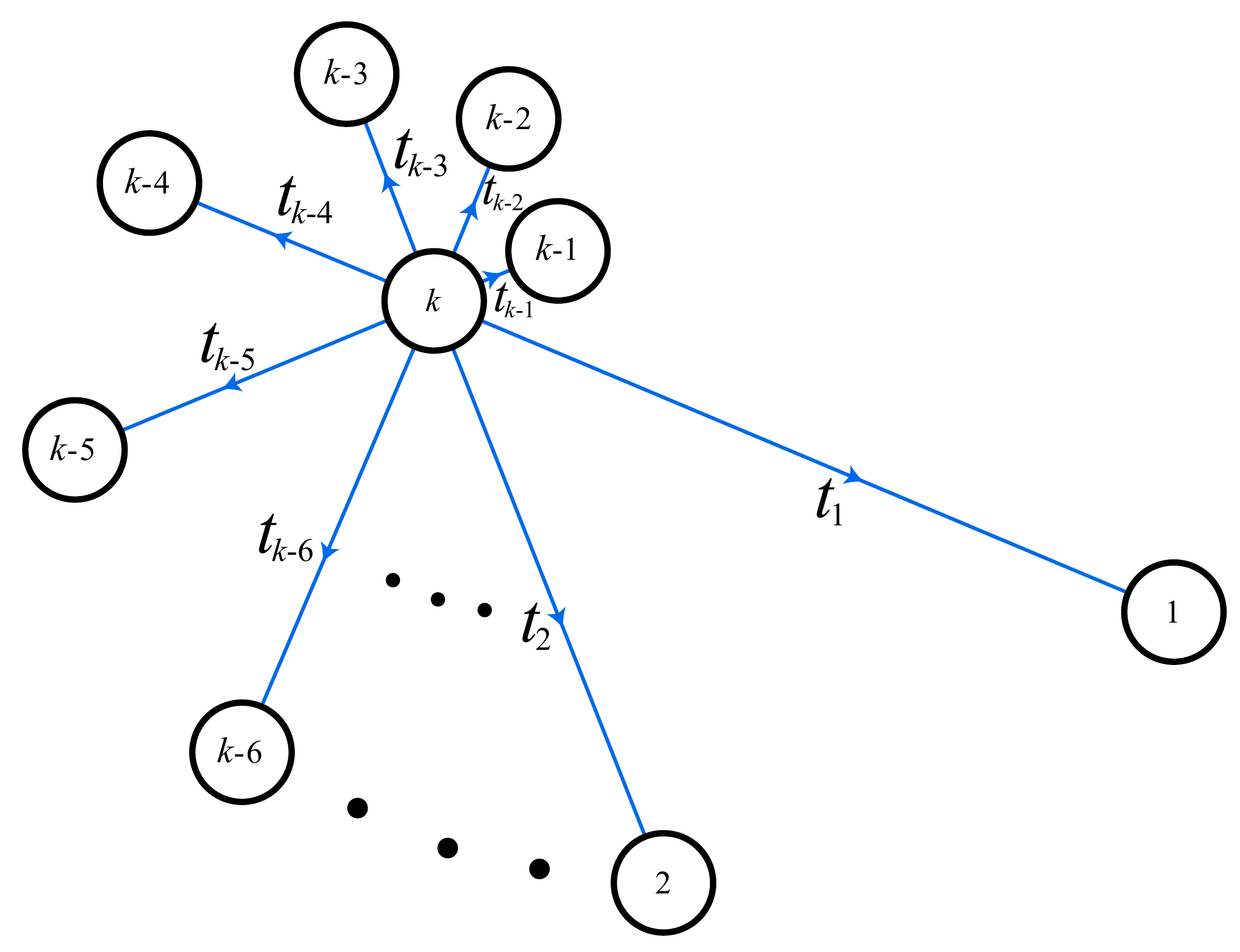}
\caption{\label{fig:flow} A schematic diagram of the interpolating flow   as a pattern of operator deformations. A circle with $i$ in it is the $i$th minimal model. Here, $k$ should be understood to be taken to infinity. There are $k{-}1$ operators in the model, labelled ${\cal O}_l$, and deformation with coefficient $t_l$ is equivalent to turning on the $l$th model. The length of the bonds/arrows signifies the differing strengths of the $t_l$.}
\end{figure}
It is natural therefore, to think of the result for the pattern of interpolating flows in equation~(\ref{eq:tk-recipe-JT}) (and in equation~(\ref{eq:tk-recipe-SJT})  to be derived shortly  for the supergravity in terms of type~0A minimal models) as simply a  specific operator deformation, or RG flow from the (infinite) $k$th model to the other models at lower~$k$. (See figure~\ref{fig:flow} for a schematic diagram.) In this sense, the suggestions  (on the one hand) of ref.~\cite{Saad:2019lba} that JT gravity is an infinite $k$ limit of the $k$th minimal model  and (on the other hand) of ref.~\cite{Okuyama:2019xbv} that JT gravity (or supergravity, as discussed in this paper) is an infinite sum of all of them, are in fact complementary. 

Next, it is important to consider low energy physics, where the features of the potential $u(x)$ at intermediate~$x$ and indeed $x{>}0$ become very important. In the case of minimal models derived from Hermitian matrix models the leading potential~$u_0(x)$ is of the form $x^{1/k}$ as $x{\to}{+}\infty$, and this leads to non--perturbative problems at low energy. In fact, ref.~\cite{Johnson:2019eik} used different minimal models that have the same large $E$ physics as those minimal models, but with different, improved,  low $E$ physics.
In those models, the leading order low energy physics of interest is simply $u_0(x){=}0$  for all $k$, a striking universal feature. See figure~\ref{fig:leading-potential}. (This behaviour is not arbitrarily imposed, but actually a solution of the underlying matrix model, as explained in  section~\ref{sec:string-equations}.)  The feature of the spectral density that results from this  can be seen by converting equation~(\ref{eq:spectral-deconstruction}) back into an $x$ integral, {\it viz.}:
\be
\label{eq:leading-low-energy-spectral}
\rhoo(E)=\frac{1}{2\pi\hbar}\int_{-|x_0|}^{\mu}\frac{dx}{\sqrt{E-u_0(x)}}\ ,
\ee
(where the lower limit $x{=}{-}|x_0|$ is where the square root vanishes) and so  the low energy regime where $u_0(x)=0$ yields the leading behaviour:
\be
\label{eq:leading-low-energy-spectral2}
\rhoo(E)=\frac{1}{2\pi\hbar}\int_0^{{\mu}}\frac{dx}{\sqrt{E}} =\frac{1}{2\pi\hbar}\frac{\mu}{\sqrt{E}}+\cdots
\ee
The choice ${\mu}{=}2$ will be made here, a convention that reproduces the leading term of equation~(\ref{eq:spectral-SJT}). Beyond low~$E$, other physics takes over and the $\mu/\sqrt{E}$ behaviour is modified, ultimately crossing over into behaviour that is described by whichever combination of the $E^{k-\frac12}$ physics  appears at high $E$. In ref.~\cite{Johnson:2019eik}, that combination was chosen to be the same as the ordinary JT gravity model, with $\mu{=}0$, giving  the same large~$E$ perturbative behaviour as the  JT matrix model  of ref.~\cite{Saad:2019lba}, but better, stable, non--perturbative physics at low $E$.

However it is possible to instead reproduce  the JT supergravity disc spectral density given in equation~(\ref{eq:spectral-SJT}). That it is also a combination of minimal models (but with different coefficients)  should follow from again expanding, and again using the integral representation~(\ref{eq:spectral-deconstruction}) of positive half--integer powers of $E$. Doing so yields: 

\bea
\label{eq:eq:SJT-deconstructed}
 \rhoo(E)&=&\frac{1}{\pi\hbar\sqrt{E}}+\frac{2}{\hbar}\sum_{k=1}^\infty
  \frac{2^{2k-1}\pi^{2k-1}}{(2k)!}E^{k-\frac12} \qquad ({\rm SJT})\nonumber\\
 &=&\frac{1}{\pi \hbar\sqrt{E}}+\frac{2}{\hbar}\int_0^E  \sum_{k=1}^\infty \frac{2^{2k-1}\pi^{2k-1}}{(2k)!}\frac{ku_0^{k-1} }{C_{k-1}}  \frac{du_0}{\sqrt{E-u_0}} \nonumber\\
 &=&\frac{1}{\pi \hbar \sqrt{E}}+\frac{2\pi}{\hbar} \int_0^E  \sum_{k=0}^\infty \frac{(\pi\sqrt{u_0})^{2k} }{(2k+2)k!^2} \frac{du_0}{\sqrt{E-u_0}} \nonumber\\
 &=&\frac{1}{\pi \hbar\sqrt{E}}+ \frac{\pi}{\hbar} \int_0^E   \frac{I_1(2\pi\sqrt{u_0})}{2\pi\sqrt{u_0}}\frac{du_0}{\sqrt{E-u_0}} \ .
 \eea
So  remarkably, for the positive powers of $E$ in this  supergravity case, a natural generalization of what was  seen for the ordinary JT  case emerges:   $f(u_0){=}\pi I_1(2\pi\sqrt{u_0})/(2\pi\sqrt{u_0})$, where $I_1(s)$ is the first modified Bessel function of the first kind ({\it c.f.} the last line in equation~(\ref{eq:JT-deconstructed})). The special~$E^{-\frac12}$ part of the spectral density will be reproduced if there is a leading contribution to the potential  of form $u_0{=}0$, as discussed above. As already mentioned, the minimal models used below for the full (not just leading order) potential will naturally have this behaviour built in. 
So this shows that the super--JT spectral density can be built out of a combination of minimal models with:
\be
\label{eq:tk-recipe-SJT}
t_k=\frac{\pi^{2k-2}}{(k!)^2}\ .
\ee

A summary of the overall picture that contrasts the JT and SJT cases is as follows. Writing $y\equiv\pi\sqrt{E}$:
\begin{eqnarray}
 {\rm SJT:} \quad{\hbar} \rhoo(E)&=&\frac{\cosh(2y)}{y} = \frac{1}{y}+\frac{2\sinh^2y}{y}\nonumber \\
&=&\frac{1}{\pi\sqrt{E}}+\pi \int_0^E\frac{I_1(2\pi\sqrt{u_0})}{2\pi\sqrt{u_0}}\frac{du_0}{\sqrt{E-u_0}} \ .\nonumber\\ 
{\rm JT:} \quad{4\pi^2}\hbar\rhoo(E)&=&{\sinh(2y)}\nonumber\\
&=& \pi \int_0^EI_0(2\pi\sqrt{u_0})\frac{du_0}{\sqrt{E-u_0}} \ . 
\end{eqnarray}

The next step is to consider the perturbative physics beyond the disc and of course non--perturbative aspects. 

\section{String Equations}
\label{sec:string-equations}

As stated before, in the approach to double--scaled matrix models that will mostly be taken here, the focus is  on the potential $u(x)$ of the associated Hamiltonian problem~(\ref{eq:schrodinger}). In the minimal string approach of old, the non--linear ODE that defines it was often called a ``string equation". A sort of master string equation that contains all the behaviour of  current interest is~\cite{Morris:1990bw,Dalley:1992qg,Dalley:1992vr,Dalley:1992br}: 
\be
\label{eq:string-equation-2}
u{\cal R}^2-\frac{\hbar^2}{2}{\cal R}{\cal R}^{\prime\prime}+\frac{\hbar^2}{4}({\cal R}^\prime)^2=\hbar^2\Gamma^2\ ,
\ee
where for the $k$th model,  
\be
{\cal R} \equiv {\tilde R}_k[u] + x\ .
\ee
Here, ${\tilde R}_k[u]$ is the  $k$th order polynomial in $u(x)$ and its $x$--derivatives defined by Gel'fand and Dikii~\cite{Gelfand:1975rn}, but normalized so that the coefficient of $u^k$ is unity. For example:
\bea
\label{eq:gelfand-dikii}
{\tilde R}_1[u]&=&u\nonumber\\
{\tilde R}_2[u]&=&u^2-\frac{1}{3}u^{''}\nonumber\\
{\tilde R}_3[u]&=&u^3-\frac12(u^{'})^2-uu^{''}+\frac{1}{10}u^{''''} \nonumber\\
&\vdots&\nonumber\\
{\tilde R}_k[u]&=&u^k+\cdots+\#u^{(2k-2)}\ .
\eea
Here a prime denotes an $x$--derivative times a factor of~$\hbar$, and in the last line the superscript $(2k{-}2)$ means that number of primes.  

String equation~(\ref{eq:string-equation-2}) was first derived (initially with~$\Gamma{=}0$) by taking~\cite{Morris:1990bw,Dalley:1992qg,Dalley:1992vr} the double--scaling limit of models of a random complex matrix $M$, with polynomial potential $V(M^\dagger M)$. Diagonalization to work in terms of the positive quantities $\lambda_i^2$, where $\lambda_i$ are eigenvalues of $M$, shows that they  are in the $(1,2)$ Altland--Zirnbauer class of matrix ensembles. It was soon realized~\cite{Dalley:1992br} that non--zero $\Gamma$ could be introduced naturally from a number of different perspectives, including one where $\Gamma$ corresponds to having added~$\Gamma$ quark flavours, or in modern language, $\Gamma$ background D--branes. There is an extra logarithmic term in the potential with coefficient $\Gamma$, which amounts to studying the $(2\Gamma{+}1,2)$ class of matrix ensembles. Later, ref.~\cite{Klebanov:2003wg} supplied a type~0A interpretation, including an understanding of~$\Gamma$ as also counting units of R--R flux. 

In almost all  the work in the literature on  equation~(\ref{eq:string-equation-2}), integer $\Gamma$ was considered the most physical choice, although it was noticed that half integer cases possessed certain interesting (but not fully explained)  properties. (Ref.~\cite{Johnson:2006ux} explored many of these properties,  suggesting a partial physical understanding in terms of minimal string theories with no background  D--branes.) {\it In this paper it will be made clear that solutions with half--integer~$\Gamma$ are physical and play a very important role.} 

For orientation, the natural  next step  is  to  study some special cases.


\subsection{Bosonic Minimal Models}
\label{sec:bosonic-minimal-models}
Consider first the case of $\Gamma{=}0$. An obvious non--trivial solution to the string equation~(\ref{eq:string-equation-2})  is ${\cal R}{=}0$, defining a subset of equations for any $k$. These are simply the ODEs defining the original $(2,2k{-}1)$ bosonic minimal models arising from double--scaling Hermitian matrix models~\cite{Brezin:1990rb,Douglas:1990ve,Gross:1990vs,Gross:1990aw}. The case $k{=}1$ is the Airy model $u(x){=}{-}x$,  the case $k{=}2$ is the Painlev\'e~I equation defining pure gravity (with the leading $u_0{=}(-x)^{1/2}+\cdots$), and $k{=}3$ is the gravitating Lee--Yang model (with the leading $u_0{=}(-x)^{1/3}+\cdots$), {\it etc.} Not much will be said about these beyond this point, in this paper, but it is perhaps useful to signpost them for orientation.

\subsection{A Universal Model}
\label{sec:bessel-and-resolvent}
Another interesting special solution is the case where ${\cal R}{=}x$, in which case there is an exact solution of the equation~(\ref{eq:string-equation-2}):
\be
\label{eq:bessel-potential}
u(x)=0+\hbar^2\frac{\left(\Gamma^2-\frac14\right)}{x^2}\ .
\ee
So the leading piece of the potential is $u_0{=}0$.  This, as we have seen, will produce a $1/(\pi\hbar\sqrt{E})$ behaviour in the spectral density $\rho_0(E)$, and the order $\hbar^2$ piece that comes next generates corrections.\footnote{This curious exact solution for $u(x)$, for $\Gamma{=}0$, was first noticed in ref.~\cite{Dalley:1992br} and referred to as the $k{=}0$  solution. It was generalized to an interesting infinite family of rational solutions in ref.~\cite{Johnson:2006ux}.} In fact, there is a nice way to organize the higher order corrections, giving an opportunity to introduce  a technique that will become useful later on. 

\subsection{A Resolvent Method}
 \label{sec:resolvent}
 In fact, the spectral density can be written as:
\be
\label{eq:density-resolvent}
\rho(E)=\frac{\rm Im}{\pi\hbar}\int_a^b {\widehat R}(x,E) dx\ ,
\ee
where the quantity ${\widehat R}(x,E){\equiv}<\!\!x|({\cal H}{-}E)^{-1}|x\!\!>$ is the (diagonal of the) resolvent of the Schr\"odinger Hamiltonian~${\cal H}$ given in equation~(\ref{eq:schrodinger}). The interval $[a,b]$ that is chosen will depend upon which $x$ regime is being studied, and will be discussed shortly. (The ``Im'' part of the prescription above means taking the imaginary part of the integrated resolvent as it approaches the real positive $E$ line.)

It is very important to note that this is {\it not} the resolvent expectation value $R(x)$ used in refs.~\cite{Saad:2019lba,Stanford:2019vob}. That is the resolvent of the raw random matrix of the matrix model, from which physics is subsequently extracted in the scaling limit. The resolvent here is that of the doubled scaled Hamiltonian that arises after double scaling. The hat will hopefully go some way to helping the reader separate the two objects when they consult those papers. The relation between them is that the $x$--integral (between appropriately chosen limits $a$ and $b$) of the resolvent of this paper is proportional to  (the scaling part of) the resolvent of those papers.

Rather usefully, ${\widehat R}(x,E)$ satisfies the Gel'fand--Dikii equation~\cite{Gelfand:1975rn}:
\begin{equation}
\label{eq:gelfand-dikii}
4(u-E){\widehat R}^2-2\hbar^2{\widehat R}{\widehat R}^{\prime\prime}+\hbar^2({\widehat R}^\prime)^2 = 1\ ,
\end{equation}
where $u{=}u(x)$, and a prime denotes a differentiation with respect to~$x$.  This equation supplies, for an input potential $u(x)$, the full (perturbative and non--perturbative) spectral density $\rho(E)$ {\it via} equation~(\ref{eq:density-resolvent}). This alternative approach for computing $\rho(E)$ is rather useful (see ref.~\cite{Johnson:2004ut} for an earlier study in this type~0A context), and in fact ref.~\cite{Johnson:2019eik} used the above equations to write a differential equation directly for $\rho(E,x)$. However, working directly with the resolvent is illuminating: Starting with $u(x){=}u_0(x)$, the leading piece in the limit $\hbar{=}0$, all derivative terms can be dropped in equation~(\ref{eq:gelfand-dikii}) and  the result~(\ref{eq:leading-low-energy-spectral}) comes from the solution  ${\widehat R}{=}{-}1/(2\sqrt{u_0(x){-}E})$ (the sign is chosen to give a positive density). The case $u_0(x){=}0$ in the integrand is between $x{=}0$ and $x{=}\mu$, giving    term ${\widehat R}{=}{-}1/(2\sqrt{-E})$ and it leads to the contribution in equation~(\ref{eq:leading-low-energy-spectral2}).  Expanding around this  leading result  gives, using equation~(\ref{eq:bessel-potential}), to  order~$\hbar^2$:
\be
{\widehat R}=-\frac{1}{2\sqrt{-E}}-\frac{\hbar^2(\Gamma^2-\frac14)}{4x^2(-E)^{3/2}}+\cdots
\ee
which yields, using equation~(\ref{eq:density-resolvent}):
\be
\label{eq:bessel-expand}
\rho(E)=\frac{1}{\pi\hbar\sqrt{E}}-\frac18\left(\Gamma^2-\frac14\right)\frac{\hbar}{\pi E^{3/2}}+\cdots
\ee
The second term should be thought of as a combination of $\rho_{1,0}(E)$ (genus one with one boundary) and $\rho_{0,2}$ (one boundary and two cross--caps), depending upon the interpretation of $\Gamma$.

Of course, there are non--perturbative contributions to~$\rho(E)$ that cannot be obtained by this perturbative procedure. It was shown in ref.~\cite{Carlisle:2005wa} that the wavefunction $\psi(E,x)$ of this particular  Hamiltonian problem (defined by potential~(\ref{eq:bessel-potential})) can be written in closed form in terms of Bessel functions of the first kind,~$J_\Gamma$:
\begin{equation}
\psi(E,x)=\frac{1}{\sqrt{2} \hbar} x^\frac12J_\Gamma\left(\frac{\sqrt{E}x}{\hbar}\right)\ ,
\end{equation}
 and the spectral density can be computed~\cite{Johnson:2019eik,doi:10.1063/1.530157,Tracy:1993xj}:
\bea
 \label{eq:bessel-density}
 \rho_J(E,{\mu})\! &=&\! \int_{0}^{\mu}\! |\psi(E,x)|^2 dx = \frac{1}{4E}\int_0^{\frac{E\mu^2}{\hbar^2}}\!\!\! J^2_\Gamma(\sqrt{t}) dt \nonumber\\
&=& \frac{{\mu}^2}{4\hbar^2}\!\left[J_\Gamma^2(\xi)\!+\!J_{\Gamma+1}^2(\xi)\!-\!\frac{2\Gamma}{\xi}J_\Gamma(\xi)J_{\Gamma+1}(\xi)\right],
\nonumber \\
&&\mbox{ where} \quad \xi\equiv{\mu\sqrt{E}}/{\hbar} \ .
 \eea
Expanding this result, with $\mu{=}2$,  yields the previously obtained perturbative terms, as well as non--perturbative terms oscillatory in $\sqrt{E}/\hbar$.

\begin{figure}[h]
\centering
\includegraphics[width=0.4\textwidth]{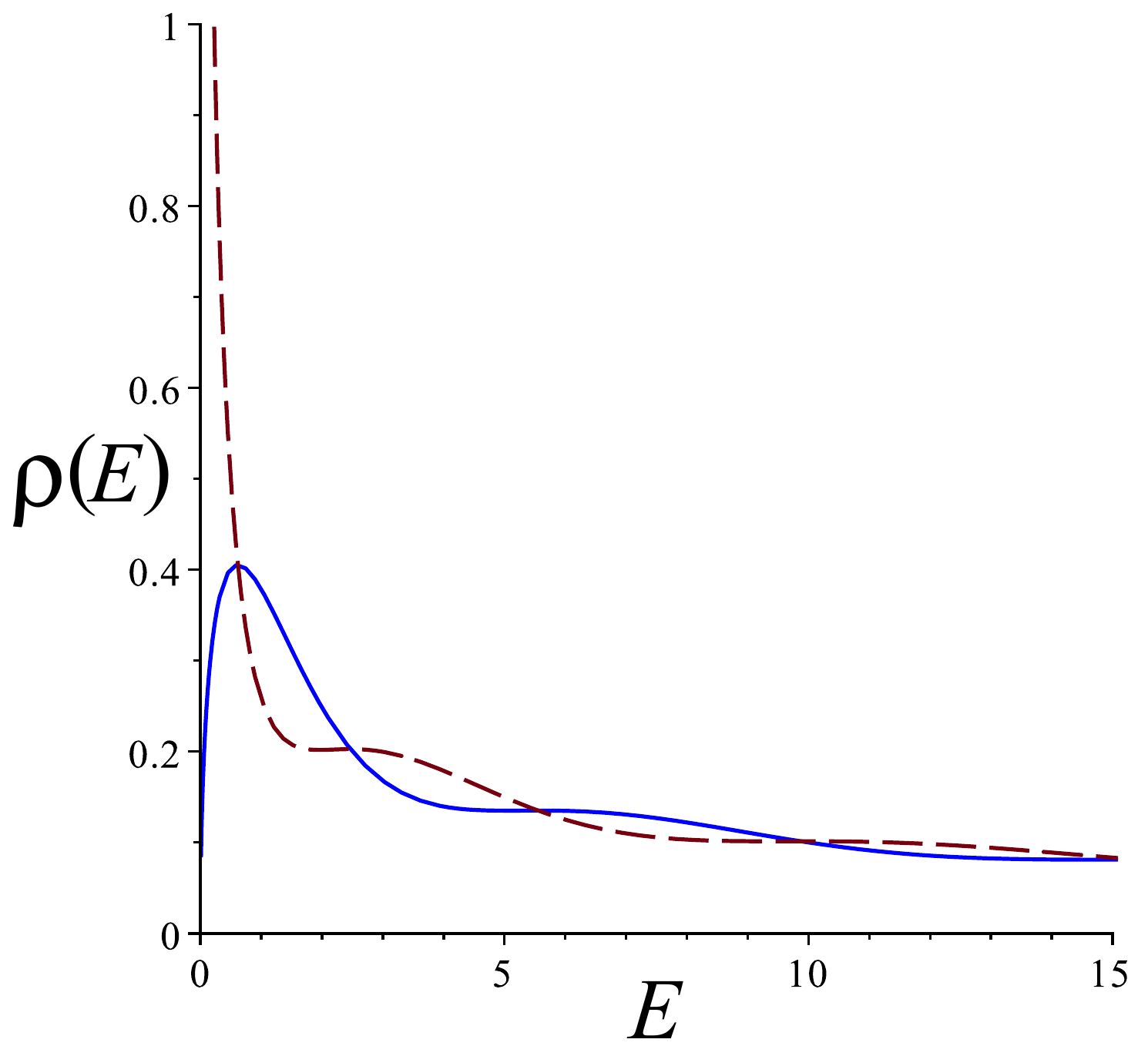}
\caption{\label{fig:special-densities} The special Bessel spectral densities  for\\ \hskip1.5cm $\Gamma{=}\frac12$ (solid) and $\Gamma{=}{-}\frac12$ (dashed).}
\end{figure}

This ``Bessel'' model is important since it furnishes an exact  model of the low energy sector for all the minimal models, and for the full JT super--gravity. For the special cases of $\Gamma{=}{\pm}\frac12$ (corresponding to the special $(0,2)$ and $(2,2)$ cases discussed in ref.~\cite{Stanford:2019vob}), things become become extremely simple. 
Since ordinary Bessel functions of half--integer order can be written in terms of trigonometric functions, specifically for this case:
  \be
  J_\frac12(\xi)=\sqrt{\frac{2}{\pi \xi}}\sin \xi\ ,\quad {\rm and} \quad J_{-\frac12}(\xi)=\sqrt{\frac{2}{\pi \xi}}\cos \xi\ ,
   \ee 
   and using $J_{\frac32}(\xi)={\xi}^{-1}J_\frac12(\xi){-}J_{-\frac12}(\xi)$,
   the asymptotic form given in  equation~(\ref{eq:bessel-expand}) truncates to the leading piece, and remarkably the full perturbative and non--perturbative parts are contained in this compact form:
   \be
   \label{eq:exact-density}
   \rho_{\pm\frac12}=\frac{1}{\pi\hbar\sqrt{E}}\mp\frac{1}{4\pi E}\sin\left(\frac{4\sqrt{E}}{\hbar}\right)\ .
   \ee
   These cases are plotted in figure~\ref{fig:special-densities}.   
   
   The $E{\rightarrow}0$ behaviour of these two cases are markedly different. For $\Gamma{=}{+}\frac12$, in the limit there is a nice cancellation between the first and second terms,  and so $\rho_{\frac12}{\to}0$.    There's a doubling instead of a cancellation for $\Gamma{=}{-}\frac12$, resulting in an $E^{-1/2}$ divergence 
for $\rho_{-\frac12}$, even after non--perturbative effects are taken into account. These two behaviours will be seen again in the full non--perturbative definition building JT supergravity out of minimal models, proposed and explored in section~\ref{sec:SJT-from-minimal}.

From the point of view of the recursive construction of Stanford and Witten~\cite{Stanford:2019vob}, the vanishing of all orders in perturbation theory (beyond the disc for the spectral density, or the disc+crosscap for the accompanying resolvent) is a sort of perturbative miracle\footnote{Also, it accompanies the fact that in this special case the whole perturbative analysis is saved from being afflicted by a divergence in the volume of the one--crosscap moduli space that would propagate to higher orders through recursion relations.}. Here, working directly with the potential $u(x)$, it is easier to isolate its origins, and to anticipate what can happen in the full model defined in the next section. Looking at the resolvent's differential equation~(\ref{eq:gelfand-dikii}), it is clear that exact vanishing of the potential at $\Gamma{=}{\pm}\frac12$ guarantees no perturbative corrections to ${\widehat R}(E,x)$ (and hence $\rho(E)$) beyond the disc. While
the potential vanishes,  there's still however a non--trivial equation for ${\widehat R}(E,x)$, and this will encapsulate the non--perturbative physics. A nice way to write it is to take another derivative. The structure of the equation is such that there is a cancellation between terms, and an overall factor of ${\widehat R}$ can be divided out, and for vanishing $u$:
\be
{\widehat R}^{'''}=-4E{\widehat R}^{'}\ ,
\ee
with an obvious solution of a constant (already found to be ${-}1/(2\sqrt{-E})$) plus an oscillatory piece with frequency $2\sqrt{E}/\hbar$. A  natural choice of its coefficient is to make the resolvent vanish at $x{=}0$, yielding: 
\be
{\widehat R}(E,x)=-\frac{1}{2\sqrt{-E}}\left[1-\exp\left(i\frac{2\sqrt{E}}{\hbar}x\right)\right]\ .
\ee
Integrating this between $x{=}0$ and $x{=}\mu{=}2$ gives  an imaginary piece which ({\it via} equation~(\ref{eq:density-resolvent}))  gives  the exact density in equation~(\ref{eq:exact-density})  for  $\Gamma{=}{+}\frac12$. (The case $\Gamma{=}{-}\frac12$  giving  the relative minus sign does not seem as natural here.)    

Crucially, there's also a  non--oscillatory term, $1/(4\pi E)$ coming from the upper limit of the integration of the exponential. It has the interpretation as a crosscap contribution, but does not appear in the density since it is real. This is the finite crosscap term discussed in ref.~\cite{Stanford:2019vob} in this $\Gamma{=}\frac12$ case (and its $\Gamma{=}{-}\frac12$ counterpart). 
   
\subsection{Type~0A Minimal Models}
\label{sec:type-0A}
\begin{figure}[h]
\centering
\includegraphics[width=0.5\textwidth]{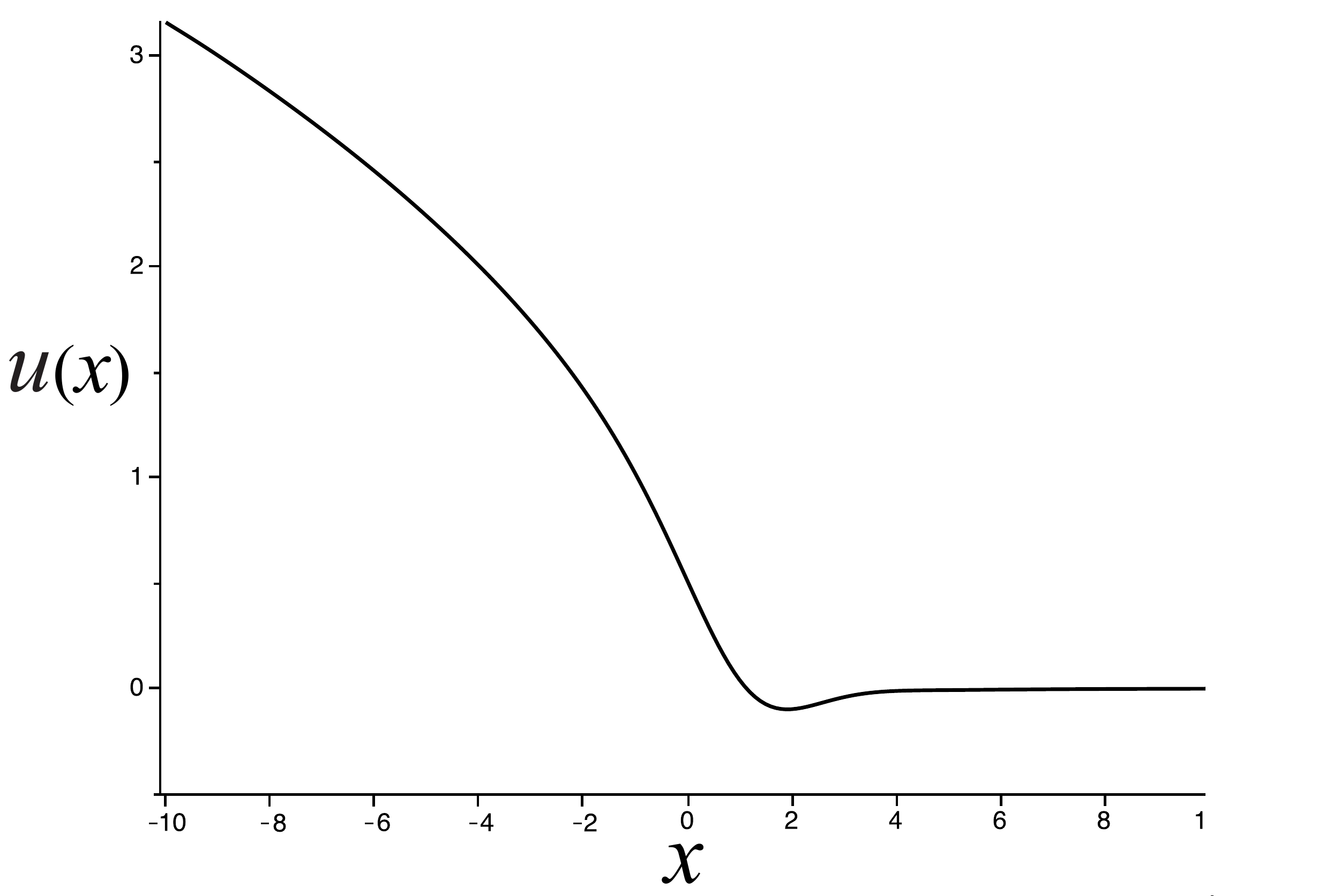}
\caption{\label{fig:potential2} The potential $u(x)$ that is supplied by equation~(\ref{eq:string-equation-2}) for the case $k{=}2$. {\it c.f.,}  figure~\ref{fig:leading-potential} for the leading part, $u_0(x)$, in this case.} 
\end{figure}

As should be clear from the previous section,   the potentials that are needed to construct  JT supergravity are ones which contain {\it both} types of behaviour, where, for  non--zero $\Gamma$:
\bea
\label{eq:leading-form-for-k}
u(x)&=&(-x)^{\frac1k}+\frac{\hbar\Gamma}{k(-x)^{1-\frac{1}{2k}}}+\cdots \qquad {x\to-\infty} \nonumber\\
u(x)&=&0+\frac{\hbar^2\left(\Gamma^2-\frac14\right)}{x^2}+\cdots \qquad {x\to+\infty}
\eea  
For $\Gamma{=}0$, the large negative $x$ behaviour is identical to that of the bosonic minimal models, derived from Hermitian matrix models. This is what inspired their original study~\cite{Morris:1990bw,Dalley:1992qg,Dalley:1992vr} as alternative formulations of minimal string theories that had the same perturbative physics as the bosonic case, but better non--perturbative behaviour. This also motivated a definition~\cite{Johnson:2019eik} of non--perturbative JT gravity with them based on the choice of $t_k$ that yields the disc spectral density~(\ref{eq:spectral-JT}) at large $E$. Note that the leading large positive~$x$ behaviour is $k$--independent, showing a kind of universality.  Figure~\ref{fig:potential2} shows the $k{=}2$ example, constructed numerically, of one of these solutions. (Compare it to  the leading piece, $u_0(x)$ shown in figure~\ref{fig:leading-potential}.)

 The $k$th model is in fact the $(2,4k)$ type~0A minimal string theory: For non--zero $\Gamma$, in any given minimal model, the first subleading term in the large $-x$ regime has the interpretation (where two integrations of $u(x)$ gives the minimal string free energy) as a disc term, with~$\Gamma$ counting the Chan--Paton labels for open string sectors, {\it i.e.,} D--branes~\cite{Dalley:1992br}.   Also in the minimal string picture, the large $+x$ regime  is a purely closed string expansion~\cite{Dalley:1992qg,Dalley:1992vr}  with the interpretation~\cite{Klebanov:2003wg} that~$\Gamma$ counts R--R flux insertions in (type~0A) string theory. In the large ${+}x$ regime, there are features of the perturbative expansion that are common to all $k$, and these will govern key aspects of the JT gravity defined by combining them, as will become clear.

These types of solutions have been studied a lot, starting with refs.~\cite{Morris:1990bw,Dalley:1992qg,Dalley:1992vr,Dalley:1992yi,Dalley:1992br}, and more recently refs.~\cite{Carlisle:2005mk}. There is strong evidence that there is  a unique, smooth solution for $u(x)$ with those asymptotics for each~$k$.  This is  proven for $k{=}1$, (with $\Gamma{=}0$), since in that case there is a map from the string equation to the Painleve~II equation, for which the relevant solution was constructed by Hastings and McLeod~\cite{Hastings1980}. A generalization of their methods was applied to the wider class of solutions in ref.~\cite{Johnson:1992pu}, providing strong evidence for their existence. Moreover, the map established in ref.~\cite{Dalley:1992br} between its solutions and those of the Painlev\'e~II hierarchy  suggests uniqueness for some non--zero $\Gamma$ {\it via} the results of refs.~\cite{Crnkovic:1990ms,Watterstam:1990qs}.
\begin{figure}[h]
\centering
\includegraphics[width=0.48\textwidth]{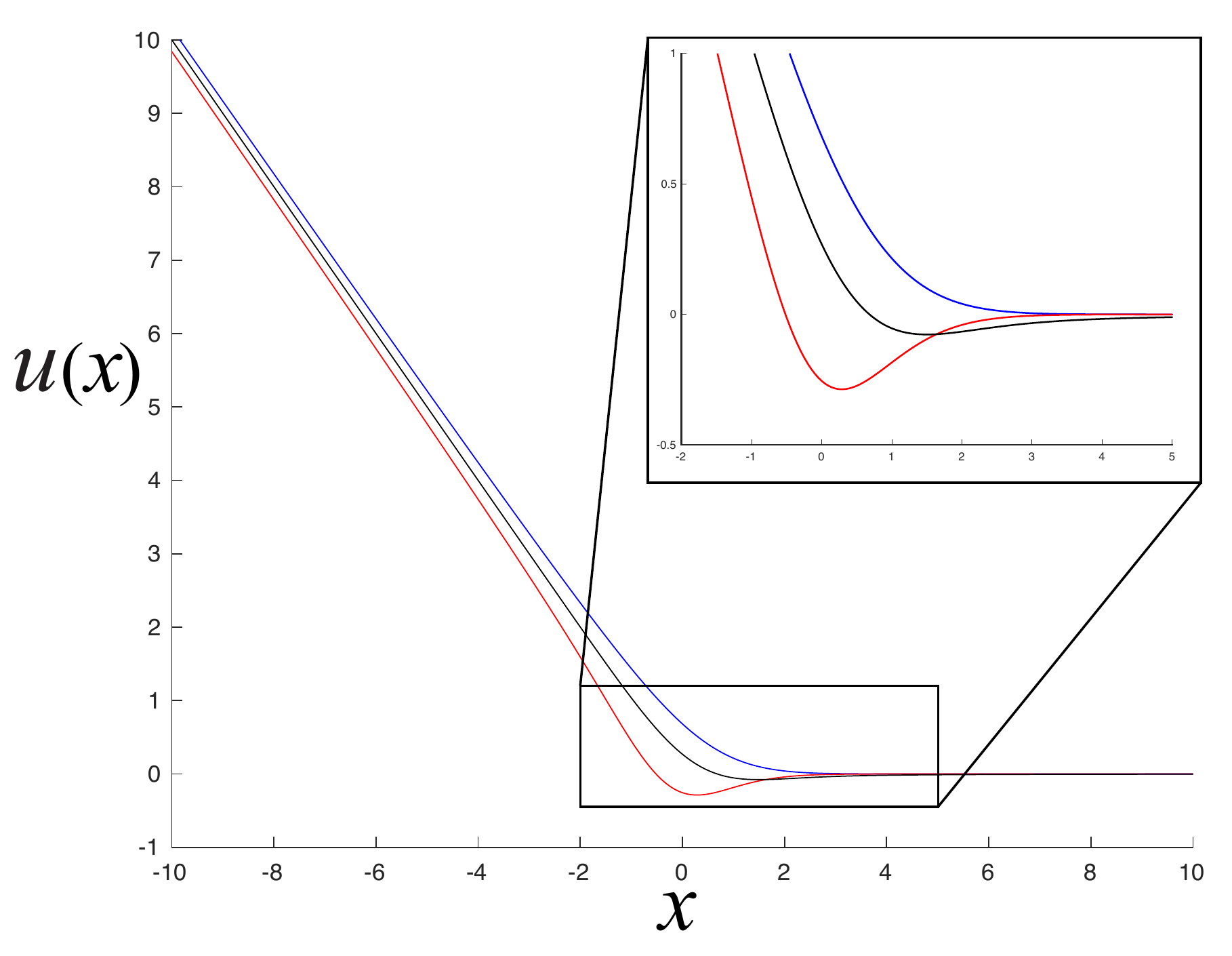}
\caption{\label{fig:potential} The (mostly) central curve is the unique $(k{=}1)$ $\Gamma{=}0$ solution to a special equation~(\ref{eq:string-equation-2}) derived from a matrix model. Also shown are the $k{=}1$ solutions for the special cases $\Gamma{=}\frac12$ (uppermost) and $\Gamma{=}\,{-}\frac12$ (lowermost)}.
\end{figure}
Moreover, the underlying  KdV flow structure (see section~\ref{sec:minimal-models}), which can evolve the unique $k{=}1$ solution into solutions for other~$k$, suggests that they exist, if not  on their own ensuring smoothness. More recently, a 't~Hooft--like  large~$\Gamma$ limit of the string equation was discussed in ref.~\cite{Klebanov:2003wg}, which also suggests smooth solutions  exist for all $k$. Numerical studies started back in ref.~\cite{Dalley:1992qg}, and since then the solutions  have been exhibited numerically for numerous~$k$ (and $\Gamma$)).  
See figure~\ref{fig:potential} for the $k{=}1$ case, with the curves for $\Gamma{=}0$, and $\Gamma{=}\pm\frac12$ superposed.

\section{
JT  supergravity\\
\hskip 0.8cm from minimal models}
\label{sec:SJT-from-minimal}
It is easy to define  a general interpolating type~0A model\footnote{Ref.~\cite{Johnson:1992pu} studied, in the context of (type~0A) minimal strings, interpolations of this sort, following similar work on 
the bosonic minimal models in ref.~\cite{Douglas:1990xv}. } as well, simply by using the string equation~(\ref{eq:string-equation-2}) with asymptotics~(\ref{eq:leading-form-for-k}), and as input:
\be
{\cal R} = \sum_{k=1}^\infty t_k {\tilde R}_k[u] + x\ ,
\ee
with the specific choice of $t_k$ given in equation~(\ref{eq:tk-recipe-SJT}), ensuring that the leading part of $u(x)$  will yield the disc contribution satisfies  equation~(\ref{eq:string-equation-sphere}), the disc spectral density will be that of the SJT model. The full string equation then gives the perturbative corrections to the disc, and the non--perturbative physics beyond. {\it In short, this is the fully non--perturbative definition of the family of JT supergravity models.}

The next two sections will examine perturbative features of this definition, coming from the two regimes where the potential (see {\it e.g.} figures~\ref{fig:potential2} and ~\ref{fig:potential}) is perturbative: large positive~$x$ and large negative~$x$. Equivalently, and perhaps more physically, these can be alternatively be thought of as regimes where~$\hbar$ is small, with $x$ either positive or negative. Perturbation theory is then organised in terms of powers of $\hbar$. Perturbative physics from positive $x$ and negative $x$ are separate families of corrections that should not be mixed together, and there is a choice as to which family should be used for expanding around the leading disc contribution given by $u_0(x)$. The JT supergravity definition has the spectral density determined by the integral~(\ref{eq:JT-definition}), with $\mu$ positive (equal to 2 in our conventions, as already discussed). Small $\hbar$ is to be understood as small compared in magnitude to $\mu$, which is in the $x{>}0$ regime. So positive $x$  is therefore  the meaningful perturbative regime. (Conversely, were $\mu$ chosen to be negative, the meaningful perturbative expansion to develop would be from the $x{<}0$ regime, but that is not the situation here.\footnote{\label{fn:hmm-remark}The case $\mu{=}0$ is special, as it resembles the choice usually made for ordinary JT gravity. To really connect to ordinary JT gravity it should be understood as $\mu{=}0^{-}$, thereby invoking the $x{<}0$ perturbation theory that coincides with the Hermitian matrix models for which the string equation is simply ${\cal R}{=}0$. This is in fact what is used in ref.~\cite{Johnson:2019eik}, to use these type 0A minimal models (with $t_k$ choice~(\ref{eq:tk-recipe-JT})) to give a non--perturbative definition of JT gravity that coincides perturbatively with that given by Hermitian matrix models.})

So perturbative corrections away from the $\rhoo(E){=}\cosh(2\pi\sqrt{E})/\pi\hbar\sqrt{E}$ already determined from the $u_0(x)$ will come from expanding $u_0(x)$ around 0 (its leading value for $x{>}0$) in the positive $x$ regime, and using the resolvent equation~(\ref{eq:gelfand-dikii}) to develop an expansion for ${\widehat R}(x,E)$, and finally using equation~(\ref{eq:density-resolvent}) with limits $[a,b]{=}[0,\mu]$ (where $\mu{=}2$ in this paper's conventions).

\subsection{Perturbation Theory: Positive $x$}
\label{sec:large-positive-x}


A most important feature  of the definition is the fact that in the  positive $x$ regime, the potential has the perturbative form $u(x){=}\hbar^2(\Gamma^2{-}\frac14)/x^2{+}\cdots$. This  controls the structure of the leading low energy behaviour, as already discussed above.
The exact model of this sector, discussed in sections~\ref{sec:bessel-and-resolvent} and~\ref{sec:resolvent}, has this leading behaviour. But the behaviour beyond this is different for each $k$. So they supply an infinite family of generalizations of the model, but with one important feature: The factor $(\Gamma^2{-}\frac14)$ multiplies the entire asymptotic series (for large positive $x$) for any $k$. For example,  for $k{=}1$: 
\bea
u(x)&=& \frac{\left(\Gamma^2-\frac14\right)\hbar^2}{x^2}-2\frac{\left(\Gamma^2-\frac14\right)\left(\Gamma^2-\frac94\right)\hbar^4}{x^5}\nonumber\\
&+&7\frac{\left(\Gamma^2-\frac14\right)\left(\Gamma^2-\frac94\right)\left(\Gamma^2-\frac{21}{4}\right)\hbar^6}{x^8}+\cdots\
\eea
and similarly for $k{=}2$:
\bea
u(x)&=& \frac{\left(\Gamma^2-\frac14\right)\hbar^2}{x^2}-2\frac{\left(\Gamma^2-\frac14\right)\left(\Gamma^2-\frac94\right)\left(\Gamma^2-\frac{25}{4}\right)\hbar^6}{x^7}\nonumber\\
&+&\frac{11}{48}\left(\Gamma^2-\frac14\right)\left(\Gamma^2-\frac94\right)\left(\Gamma^2-\frac{25}{4}\right)\times\nonumber\\
&&\hskip1.5cm\frac{(48\Gamma^4-1240\Gamma^2+8371)\hbar^{10}}{x^{12}}+\cdots
\eea
From here it is possible to work out ${\widehat R}(x,E)$ and hence the contributions this sector makes to the density $\rho(E)$ beyond the orders given in equation~(\ref{eq:bessel-expand}), using the resolvent techniques of  section~(\ref{sec:resolvent}). For example, for $k{=}1$, the next order is:
\bea
\rho_{2,0}(E)+\rho_{1,2}(E)+\rho_{0,4}(E) &=&\nonumber\\
 &&\hskip-3.5cm-\frac{\hbar^3}{128\pi} \left(\Gamma^2-\frac14\right)\left(\Gamma^2-\frac94\right)\frac{(1-E)}{E^{5/2}}\ ,
\eea
while for $k{=}2$ it is  almost the same  at this order (but with the $(1-E)$ in brackets replaced by $\frac12$), followed by:
\bea
\rho_{4,0}(E)+\rho_{2,2}(E)+\rho_{1,4}(E)+\rho_{0,6}(E) =&&\\
 &&\hskip-6.0cm-\frac{\hbar^5}{(32)^2\pi}\left(\Gamma^2-\frac14\right)\left(\Gamma^2-\frac94\right)\left(\Gamma^2-\frac{25}{4}\right)\frac{(1-\frac43 E^2)}{E^{7/2}}
\ ,\nonumber
\eea
at the next order.

Several remarks are due at this point. Perhaps the most important is that for $\Gamma{=}\pm\frac12$, the entire perturbative series beyond the disc order $\rho_{0,0}$ vanishes ---{\it for all the general $k$ models from which the model is built}--- as happens in the special case of the exact Bessel model of section~\ref{sec:resolvent}, and anticipated in Stanford and Witten's general perturbative analysis~\cite{Stanford:2019vob} for the full $(0,2)$ and $(2,2)$ supergravities. This is highly suggestive that this prescription for building JT supergravity from these minimal models is correct. 

Indeed, this feature is definitely present in the full interpolating theory that is proposed for the complete interpolating definition. This follows from the special nature of the equation in this regime. The whole perturbative solution of equation~(\ref{eq:string-equation-2}), regardless of the form of the input ${\cal R}$ (interpolating or not), is seeded by the $k$--independent leading solution $u{=}\hbar^2(\Gamma^2{-}\frac14)/x^2$. As a demonstration, pick the interpolating case 
\be
\label{eq:interpolating-example}
{\cal R}=t_2{\tilde R}_2+t_1{\tilde R}_1+x= t_2\left(-\frac13 u^{\prime\prime}+u^2\right)+t_1u+x\ .
\ee
Working perturbatively, the solution leads with the universal piece, followed by $k{=}1$ behaviour, with $k{=}2$ behaviour appearing at the next  order, then both appearing mixed in the expected non--linear fashion:
\bea
u(x)&=&\left(\Gamma^2-\frac14\right)\Biggl\{\frac{\hbar^2}{x^2}+\left(\Gamma^2-\frac94\right)\Biggl\{-2\frac{t_1\hbar^4}{x^5}\nonumber\\
&+&\frac{\hbar^6}{x^8}\left[7t_1^2\left(\Gamma^2-\frac{21}{4}\right)-2t_2x\left(\Gamma^2-\frac{25}{4}\right)\right]\nonumber\\
&+&\frac{\hbar^8}{x^{11}}\left[t_1^3({\rm poly}_1)+t_1t_2({\rm poly}_2)\right]\Biggr\}+\cdots\Biggr\}
\ ,
\eea 
where ${\rm poly}_{1,2}$ are fourth order polynomials in $\Gamma$.
The key feature is  the factor of $\Gamma^2{-}\frac14$. This has an obvious generalization to the fully interpolating model. So  the vanishing of all orders at $\Gamma{=}{\pm}\frac12$ is guaranteed in the full proposed JT supergravity, in this perturbative regime.

Another remark is about the (probably already evident)  intriguing pattern that is emerging. As $k$ increases there are further special points at half integer~$\Gamma$. In such cases, the entire series for $u(x)$ truncates to the leading $C\hbar^2/\bx^2$ term for some constant $C$.  For example,~$k{=}1$ has the case $\Gamma{=}{\pm}\frac32$, for which $u(x){=}2\hbar^2/\bx^2$ exactly. The  $k{=}2$ model also has that case, and in addition $\Gamma{=}\pm\frac52$, for which $u(x){=}6\hbar^2/\bx^2$ exactly, and so on to higher $k$. Each of these has their exact spectral density at low energy given by equation~(\ref{eq:bessel-density}), with the physics at all energies furnished by the complete string equation~(\ref{eq:string-equation-2}).  There is again a truncation of the perturbative series to a finite number of terms, and all the non--perturbative physics can be written in terms of factors of oscillatory pieces with frequency $2\sqrt{E}/\hbar$.
This follows from the fact that, just as before,  Bessel functions of half--integer order can be written as combinations of trigonometric functions. 

For example, working out the $\Gamma{=}\frac32$ case explicitly is interesting. The perturbative contributions are the disc and the term that would previously be interpreted as the disc$+$handle added to the disc and two crosscaps, given by equation~(\ref{eq:bessel-expand}) with $\Gamma{=}\frac32$ inserted. Once again there are real terms in $\int_0^2 {\widehat R}(x,E)dx$ coming from the lower limit, that correspond to crosscap order, as happened for $\Gamma{=}\frac12$. However, in this case they are divergent. It is natural to suggest that this corresponds to the crosscap divergence seen by Stanford and Witten for $\Gamma{\neq}\pm\frac12$, but this should be further explored. Nevertheless the density, which seems to involve only even numbers of crosscaps, is finite and gives a well--behaved function, presumably defining a sensible theory for these broader values of $\Gamma$.

   The presence of other special points for half--integer $\Gamma$ giving $u{\sim}\hbar^2/\bx^2$ suggests further special circumstances in the theory of JT supergravity. In fact,  there's a much richer story, which was elucidated in ref.~\cite{Johnson:2006ux}. There, it was shown that at half integer $\Gamma$ there are many {\it rational solutions} for $u(x)$ for each $k$, in the form of a ratio of polynomials in $x$ differing by two orders. This gives an expression for $u(x)$ that starts as $C\hbar^2/\bx^2+\cdots$  for some constant $C$ and then truncates at some order.

\subsection{Perturbation Theory: Negative $x$}
\label{sec:large-negative-x}

As already mentioned above, perturbation correctons beyond $\rho_0(E)$ comes from the expansion in the positive~$x$ regime discussed in  the previous section. There, it was confirmed that certain special features of JT supergravity observed by Stanford and Witten are naturally reproduced by the special properties of the string equation in this regime.  However, it is interesting to explore the structure of the negative $x$ expansion. (This would be physically relevant for the case of $\mu{=}0^{-}$, for example, relevant to ordinary JT gravity were $\Gamma{=}0$---see footnote~\ref{fn:hmm-remark}.) Moreover, it will uncover a useful  piece of physics missing from the perturbative description of the previous section.

For each $k$, the leading form is given in the first line of equation~(\ref{eq:leading-form-for-k}). As already discussed, the first (classical) piece generates the contribution to the disc order that rises as $E^{k-\frac12}$, for each~$k$.  In this regime, there is no universal leading form for~$u(x)$ across all models, and so general statements about the nature of the potential in the interpolating model are harder to make. It is useful however to look at some of the behaviour at individual~$k$, and the case of $k{=}1$ is a good starting point. Going to a few orders beyond the leading ones, the potential is:
\bea
\label{eq:k1-potential-minus-direction}
u(x) &=& -x\pm\frac{\Gamma\hbar}{(-x)^{1/2}}-\frac12\frac{\Gamma^2\hbar^2}{x^2}\pm\frac{5}{32}\frac{\Gamma(4\Gamma^2+1)\hbar^3}{(-x)^{7/2}}\nonumber\\&&\hskip2cm+\frac{1}{8}\frac{\Gamma^2(8\Gamma^2+7)\hbar^4}{x^5}\cdots
\eea
Just as before, the resolvent technique of section~\ref{sec:resolvent} can be used to study the resulting spectral density. Some care must be taken. Here, the integral over $x$ is from $-\infty$ to $0$, and as before, finite physics can come from either limit. As a test of this methodology, consider the case $\Gamma{=}0$. Then the expansion of the resolvent  is entirely generated from the leading $u(x){=}{-}x$ term, and the result for the density should coincide with the result (expanded) of the Airy model. Indeed, expanding~(\ref{eq:gelfand-dikii}) gives:
\bea
{\widehat R}(x,E)_{\Gamma{=}0}&=&-\frac{1}{2\sqrt{-(x+E)}}-\frac{5}{64}\frac{\hbar^2}{(-(x+E))^{7/2}}\nonumber\\
&&\hskip0.8cm-\frac{1155}{4096}\frac{\hbar^4}{(-(x+E))^{13/2}}+\cdots
\eea
and so integrating and using equation~(\ref{eq:density-resolvent}) gives:
\be
\label{eq:expanded-airy}
\rho(E)_{\Gamma=0}=\frac{\sqrt{E}}{\pi\hbar}+\frac{1}{32\pi}\frac{\hbar}{E^{5/2}}
-\frac{105}{2048\pi}\frac{\hbar^3}{E^{11/2}}+\cdots\ ,
\ee
which is indeed the first few terms of the large~$E$  expansion of the exact expression for spectral density of the Airy model:
\be
\rho_{\rm Ai}(E)=\hbar^{-2/3}\left({\rm Ai}^\prime(\zeta)^2-\zeta{\rm Ai}(\zeta)^2\right)\ ,\qquad \zeta\equiv -\hbar^{-2/3}E\ .
\ee
The first term in equation~(\ref{eq:expanded-airy}) is the disc, accompanying the $1/(\pi\sqrt{E})$ present in equation~(\ref{eq:bessel-expand}), the next has a handle added, then two handles, and so forth. 

Having done that, consider $\Gamma{\neq}0$. Integrating the expanded resolvent gives, at the first two non--trivial orders with non--zero $\Gamma$:
\be
\left[-\frac12\frac{\pm\Gamma\hbar\sqrt{-x}}{E\sqrt{-(x+E)}}-\frac{1}{32}\frac{4\Gamma^2\hbar^2}{(-(x+E))^{5/2}}\right]_{-\infty}^0\ ,
\ee
where a term that is either zero or divergent at the limits has been neglected. The first term produces a  crosscap term $\pm\hbar\Gamma/2E$ from the lower limit, zero from the upper, and the second term produces an imaginary piece from the upper limit which produces a double crosscap contribution to the density, and zero from the lower. Checking a few more orders in this manner completes the expanded Airy result~(\ref{eq:expanded-airy}) for the density to:
\bea
\rho(E)&=&\frac{\sqrt{E}}{\pi\hbar}+\frac{(4\Gamma^2+1)}{32\pi}\frac{\hbar}{E^{5/2}}\nonumber\\
&&\hskip0.5cm-\frac{(336\Gamma^4+664\Gamma^2+105)}{2048\pi}\frac{\hbar^3}{E^{11/2}}+\cdots
\eea
In fact, the crosscap term must be treated with care, and it produces a special contribution.\footnote{The Author thanks Felipe Rosso for pointing  out this term's contribution.} A single pole, $1/E$, produces an imaginary part discontinuity that is a delta function: $1/(E{-}i\epsilon){-}1/(E{+}i\epsilon)=2iE/(E^2{+}\epsilon^2)\to2\pi i \delta(E)$, and so there is a delta function contribution to the density from $\Gamma$ states at zero energy: $\rho(E)_{\rm crosscap}=\pm\Gamma\delta(E)$. An analogous term arises for any $k$, and presumably for the full interpolating model, but it is the $k{=}1$ model that will dominate the low energy physics of the interpolating model. 

Such a contribution, and its significance, was discussed in ref.~\cite{Stanford:2019vob} (by Laplace transform it corresponds to  adding~$|\Gamma|$ to the partition function $Z(\beta)$). Even though it was found in the ``wrong'' perturbative regime here, it is relevant to JT supergravity. Such a term is invisible in $x{>}0$ perturbation theory and should be accessible only using a non--perturbative approach. In a sense, the $x{<}0$ perturbation theory of this section (relevant to a negative~$\mu$ 
theory) acts as a non-perturbative probe of the model from the point of view of the positive~$\mu$ case in hand. The result can be trusted because it is  $\mu$--independent.

\subsection{Non--Perturbative Results}
\label{sec:non-perturbative}
Now it is time to turn to non--perturbative features of these type 0A minimal models, including new ones that go  beyond the special features of the exact Bessel model of section~\ref{sec:resolvent}. They will act as toy models of the non--perturbative features of the full JT supergravity.

\begin{figure}[h]
\centering
\includegraphics[width=0.45\textwidth]{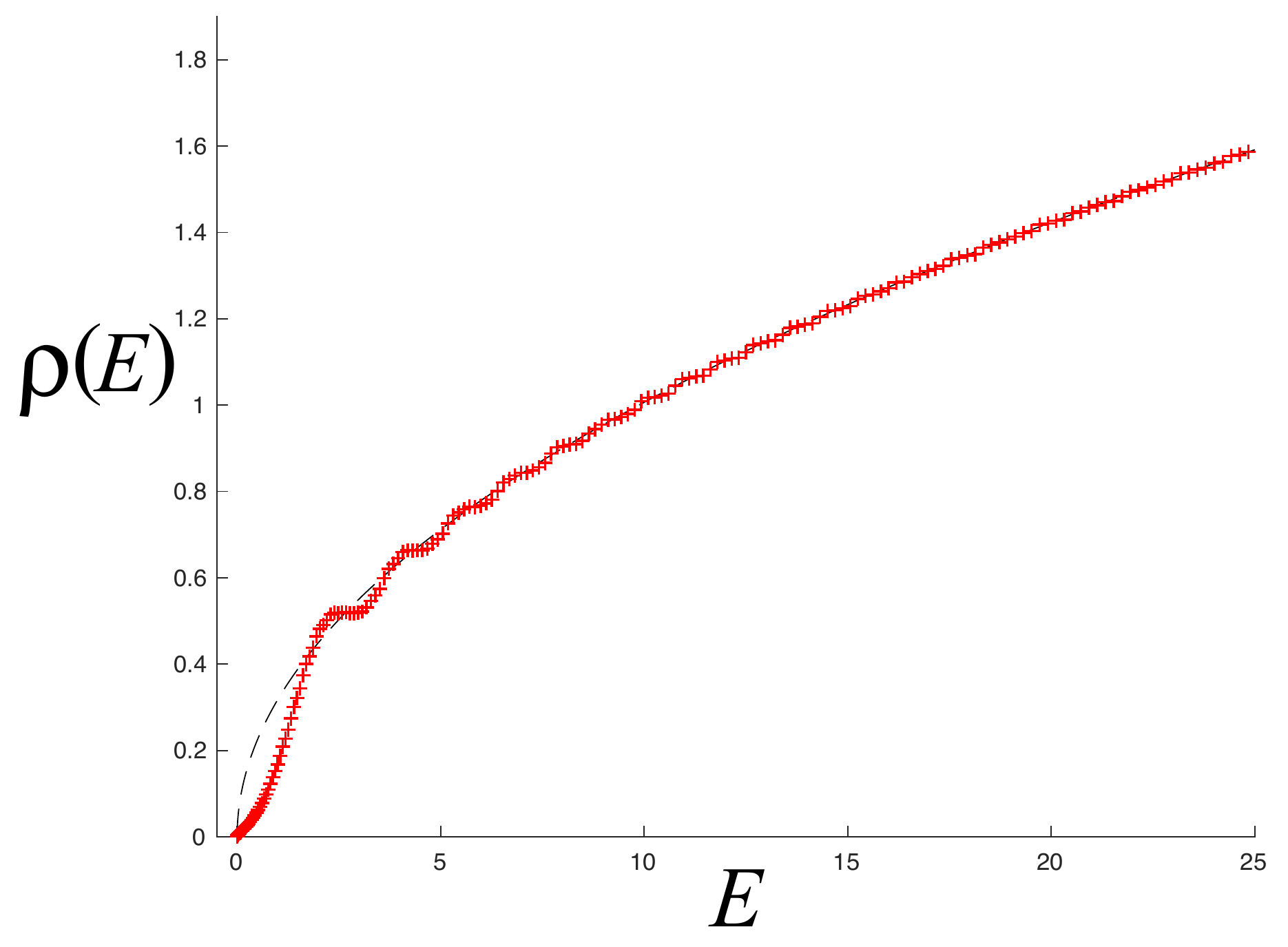}
\caption{\label{fig:full-densityA} The  $k{=}1$  spectral density for $\Gamma{=}\frac12$.}
\end{figure}

\begin{figure}[h]
\centering
\includegraphics[width=0.45\textwidth]{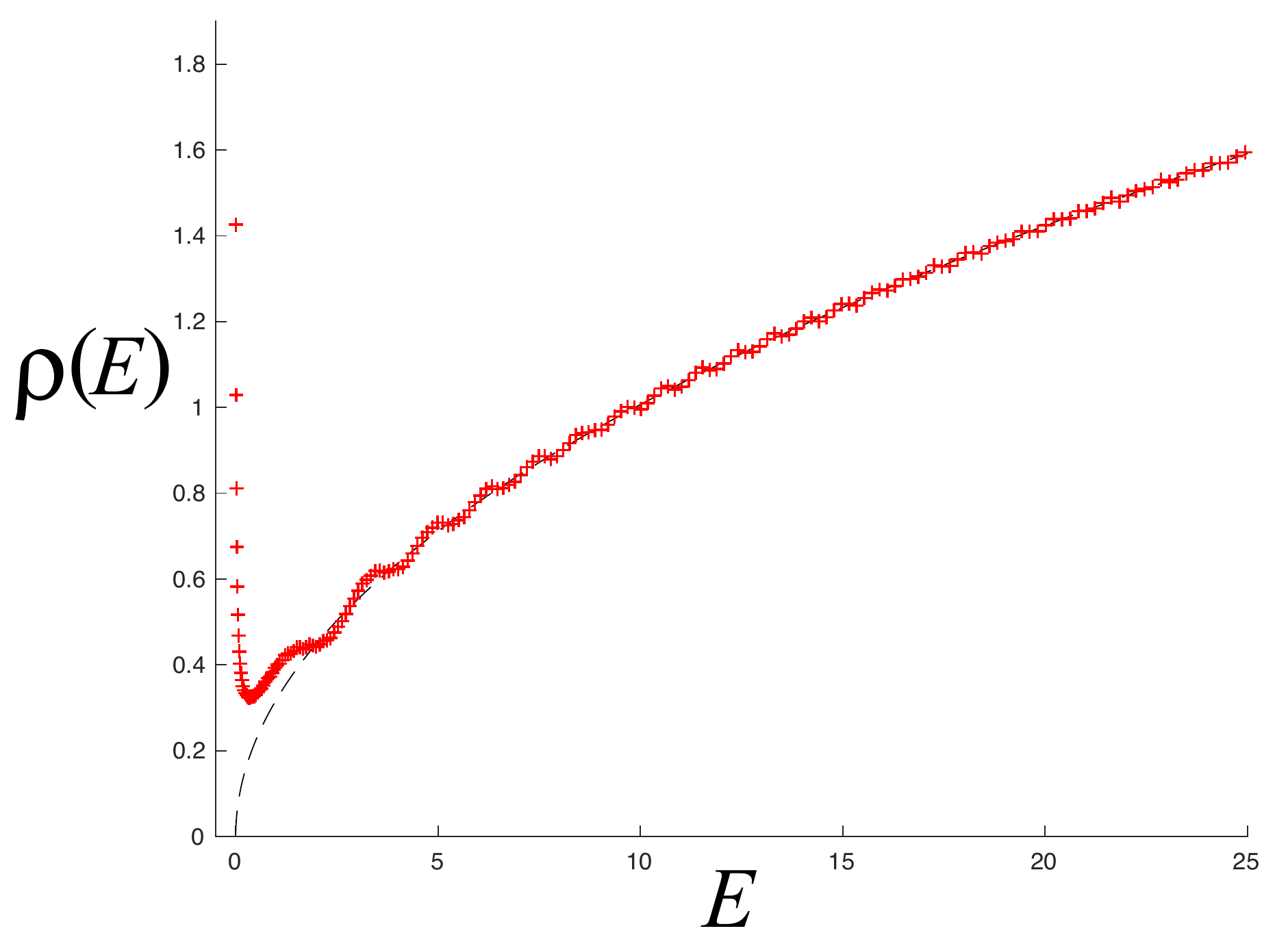}
\caption{\label{fig:full-densityB} The  $k{=}1$  spectral density for  $\Gamma{=}{-}\frac12$.}
\end{figure}

While it is not possible to write down the complete spectral density for the fully interpolating minimal models (since the string equation becomes formally of infinite order),  many of the key features are clear from looking at any particular~$k$. For $k{=}1$, for example, the spectral densities for the cases $\Gamma{=}{\pm}\frac12$ were computed numerically using the same techniques employed\footnote{Using a matrix  Numerov method~\cite{doi:10.1119/1.4748813}, the Schr\"odinger problem of equation~(\ref{eq:schrodinger}) was solved  with ${-}100{\leq} x {\leq}{+}100$ on a grid of $4000{\times}4000$. A suitable normalization was performed for the  4000 eigenfunctions  and then the spectral density was constructed using a simple trapezoidal integration. See ref.~\cite{Johnson:2019eik} for details.}  for $\Gamma{=}0$ in ref.~\cite{Johnson:2019eik}   and they are presented in figures~\ref{fig:full-densityA} and~\ref{fig:full-densityB}, respectively. 
To the left there is the $1/\sqrt{E}$ behaviour at the disc level, plus non--perturbative corrections, and just as in the special $\Gamma{=}\pm\frac12$ models studied exactly in section~\ref{sec:resolvent} (see figure~\ref{fig:special-densities}) for  $\Gamma{=}{+}\frac12$  non--perturbative effects  cancel the divergence to zero, while for $\Gamma{=}{-}\frac12$ it is enhanced. Rather than falling off at larger $E$ (as the special Bessel models do), in this case new physics turns on and generates the  rise to the right, attaining the   $E^{\frac12}$ disc asymptote at large $E$.  A final important non--perturbative feature is of course the oscillatory modulations, indicative of the underlying random matrix model structure, showing the effects of repulsion of the eigenvalues. The basic frequency is $2\sqrt{E}/\hbar$ (universal for all $k$, and hence for the full JT supergravity, because of the structure of the resolvent equation~(\ref{eq:gelfand-dikii})), increasing to the right, but the amplitude is also suppressed at larger $E$.   

Note that these  features are present for  any $k$, and the full interpolating model defining the JT supergravity. (Of course, there the rise with $E$ is not $E^{k-\frac12}$ as  for the $k$th model, but rather exponential.)

\begin{figure}[h]
\centering
\includegraphics[width=0.45\textwidth]{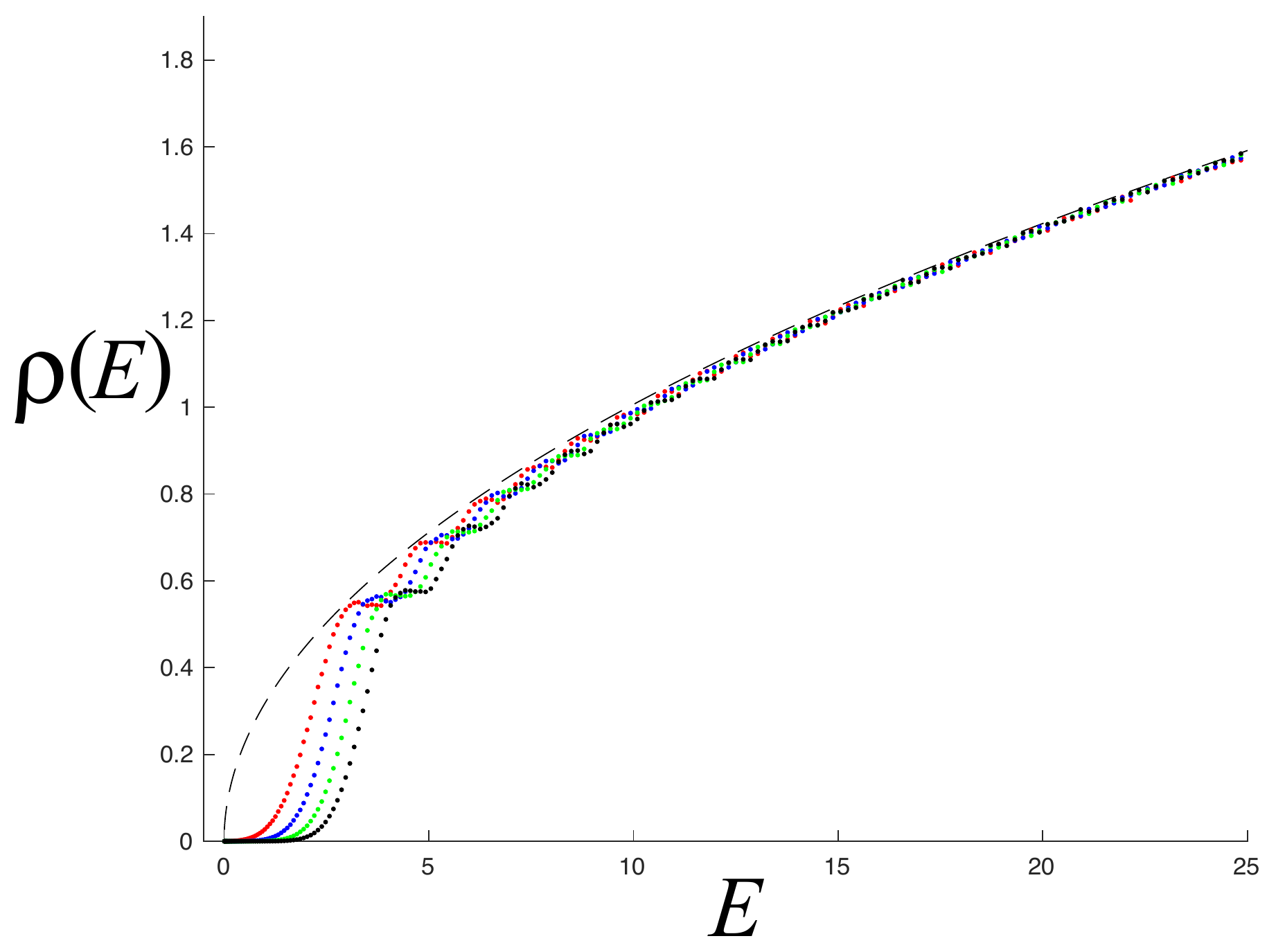}
\caption{\label{fig:full-densityC} The  $k{=}1$  spectral density for $\Gamma{=}\frac32,\frac52,\frac72$ and $\frac92$, successively, beginning from left to right.}
\end{figure}

Finally, the non--perturbative spectral density for other half--integer $\Gamma$ cases can be readily computed. Again, for $\Gamma{<}0$ the low $E$ behaviour is divergent. Figure~\ref{fig:full-densityC} shows the case of $k{=}1$ with $\Gamma{=}\frac32,\frac52,\frac72$, and $\frac92$ superimposed. 

It is likely that these cases (and the full JT supergravities built out of them) can be  generated from the $\Gamma{=}\frac12$ case by the transformation discovered in ref.~\cite{Carlisle:2005mk}, that changes $\Gamma$ by an integer {\it via} a special ``B\"acklund'' transformation. (That work also noticed several peculiar properties of $\Gamma{<}0$ that might be relevant here.)

\section{Discussion}
\label{sec:conclusions}

As models of 2D quantum gravity, minimal string theories (defined by the double--scaling limit of random matrix models~\cite{Brezin:1990rb,Douglas:1990ve,Gross:1990vs,Gross:1990aw}) produced a great deal of excitement~30 years ago because they captured, in a very compact manner, both the perturbative and non--perturbative physics of the dynamical topology of spacetime (the world--sheet of the string). The beautiful demonstrations~\cite{Saad:2019lba,Stanford:2019vob} that  JT gravity and supergravity can be written as double--scaled matrix models have renewed a lot of interest in the topological dynamics of 2D gravity, but the results (expressed through recursion relations connecting  topologies) are intrinsically perturbative  (mostly). The work presented in this paper (and a recent earlier one~\cite{Johnson:2019eik}) is based on the idea that building  JT gravity and supergravity out of minimal string models allows the powerful techniques of the older framework to be used as a complement to the recursive approach, and moreover to help define the non--perturbative sector.  (Note that a relation to minimal strings was suggested in ref.~\cite{Saad:2019lba}, and another in ref.~\cite{Okuyama:2019xbv}. Section~\ref{sec:minimal-models} of this paper argues that the two suggestions are  complementary.)

The results of this paper show that the type~0A minimal string models are the ideal components with which to build the JT supergravities that were classified in ref.~\cite{Stanford:2019vob} as being in the  $(2\Gamma{+}1,2)$ Altland--Zirnbauer class. The precise recipe for combining them was found, which yields the disc--level result from the super--Schwarzian approach. Key non--trivial features of the perturbative physics were reproduced (and in fact made manifest in the minimal string formalism). The construction provides a stable and (probably) unique  non--perturbative completion of the models, while pointing to new interesting features beyond the cases $\Gamma{=}0,{\pm}\frac12$ (see below). (A companion paper~\cite{Johnson:2020exp} explores this construction further, solving the interpolating string equation and computing several physical quantities.)

The technique used in this paper can probably be extended into a number of directions.
On the one hand, it would be interesting to formulate other quantities  (correlation functions of $Z(\beta)$, the spectral form factor, {\it etc.}) in this language, thus opening a useful new window on their physical properties. On the other hand, formulating the wider set of JT gravities and supergravities (as classified 
in ref.~\cite{Stanford:2019vob})  ought to be possible. A difficulty  might be that it is not clear if all the double--scaled matrix models of interest (and hence the minimal strings), have an  associated Hamiltonian analogous to the one discussed in this paper ({\it i.e.,} equation~(\ref{eq:schrodinger})), where $u(x)$ is supplied by a string equation. Its presence played a central simplifying role in the construction. In many cases, the analogue of~$u(x)$ is a combination of two or more functions, with coupled string equations linking them. The operator that, when double--scaled, becomes  ${\cal H}$ in one--cut Hermitian, complex, and unitary matrix model cases, does not seem to yield a suitable ${\cal H}$  in those more general cases. (See {\it e.g.,} refs.~\cite{Brezin:1990dk,Brezin:1990xr} for more on this issue of relevance to the $\beta{=}1,4$ Dyson--Wigner cases.)  Nevertheless, perhaps even in such cases a simple effective ${\cal H}$  for which the spectral problem matches that of the JT system can be found. In fact, the result  that the various random matrix ensembles each have an associated JT gravity encourages the  conjecture that {\it such an~${\cal H}$ must exist}. However there is no guarantee that it  emerges at the level of individual minimal string models. That could just be a happy circumstance in the cases discussed in this paper. 
Finally there were a number of results and observations in the body of the paper that hinted at larger structures that are worth  further investigation. For example:

$\bullet$ In section~\ref{sec:minimal-models}, in determining the combination of minimal models that yields the disc spectral density,  the 
version of  the function that went into the integral transform~(\ref{eq:spectral-deconstruction}) was a simple (and striking) generalization of that for ordinary JT gravity: $I_1(s)/s$ {\it vs.}~$I_0(s)$, where $s{=}2\pi\sqrt{u_0}$, where $u_0$ is the classical potential and~$I_n(s)$ is the $n$th modified Bessel function of~$s$. Perhaps there is a generalization ($I_n(s)/s^n$ suggests itself) that plays a role in defining other kinds of JT or JT--like systems.  

$\bullet$ This paper's definition of JT supergravity by using  component minimal models yielded (rather naturally because of the structure of the string equation) key properties of the $(0,2)$ and $(2,2)$  models that had been observed in ref.~\cite{Stanford:2019vob}.  It is clear that it also supplies a definition for a wider class of models: $(2\Gamma{+}1,2)$. It would be interesting to explore more properties of these, seeing if they are on at least equal physical footing to the cases of $\Gamma{=}{\pm}\frac12$, for example. 

The fact that other half--integer~$\Gamma$ cases can be reached by acting with the  B\"acklund transformation derived in ref.~\cite{Carlisle:2005mk}  (which explicitly gives the potential $u(x;\Gamma{\pm}1)$ if $u(x;\Gamma)$ is known) seems worth studying in this context. 
Perhaps the transformation has an interpretation as inserting the R--R punctures of ref.~\cite{Stanford:2019vob}.  A connection seems natural:  In ref.~\cite{Carlisle:2005mk} it was pointed out that (since B\"acklund transformations change a solution's soliton number, in the associated KdV context)  $\Gamma$ is to be associated with a special class of, it turned out,  zero--velocity solitons present in the associated integrable system. In addition to the observation made toward the end of subsection~\ref{sec:large-negative-x}, this  connects $\Gamma$ nicely to the index $\nu$ counting additional zero--energy states in ref.~\cite{Stanford:2019vob}.

\medskip

$\bullet$ Additionally, the fact that the string equation~(\ref{eq:string-equation-2}) has  a rich family~\cite{Johnson:2006ux} of rational function solutions for half--integer $\Gamma$ may well have an application in the study of  JT systems.

\smallskip

It is hoped that these and other issues and ideas will yield useful results to be reported soon.

\medskip
 
 \begin{acknowledgments}
CVJ  thanks Felipe Rosso for   questions and conversations,   Robie Hennigar, Krzysztof Pilch, Douglas Stanford and Andrew Svesko for  questions and remarks,  Edward Witten for  drawing his attention to issues in JT gravity nine months  ago, and all of the students in the USC High Energy Theory group for their willingness to learn a bit about ``the old ways''  of matrix models. CVJ also thanks  the  US Department of Energy for support under grant  \protect{DE-SC} 0011687, and, especially during the pandemic,  Amelia for her support and patience.    
\end{acknowledgments}

\bigskip

\bibliographystyle{apsrev4-1}
\bibliography{super_JT_gravity}

\end{document}